\documentclass[aps,prl,twocolumn,showpacs,groupedaddress]{revtex4}
\usepackage[dvips]{graphicx}
\usepackage{amssymb}
\usepackage{amsmath}
\begin{document}
\title{Integer and fractional charge Lorentzian voltage pulses analyzed in the framework of Photon-assisted Shot Noise}
\author{J. Dubois$^{1}$, T. Jullien$^{1}$, C. Grenier$^{2,3}$, P. Degiovanni$^{2}$, P. Roulleau$^{1}$, and D. C. Glattli$^{1}$}

\affiliation{\vspace{5pt} $^{1}$ CEA, SPEC, Nanoelectronics group, URA 2464,
 F-91191 Gif-Sur-Yvette, France.\\
$^{2}$ Universit\'{e} de Lyon - F\'{e}d\'{e}ration de Physique Andr\'{e} Marie Amp\`{e}re\\
CNRS - Laboratoire de Physique de l'Ecole Normale Sup\'{e}rieure de Lyon\\
46 Allée d'Italie, 69364 Lyon Cedex 07, France\\
$^3$ Centre de Physique Th\'{e}orique (CHPT), Ecole Polytechnique, 91128 Palaiseau Cedex, France}

\date{January 22, 2013}
\begin{abstract}

The periodic injection $n$ of electrons in a quantum conductor using periodic voltage pulses applied on a
contact is studied in the energy and time-domain using shot noise computation in order to make comparison with
experiments. We particularly consider the case of periodic Lorentzian voltage pulses. When carrying integer charge, they are known to
provide electronic states with a minimal number of excitations, while other type of pulses are all
accompanied by an extra neutral cloud of electron and hole excitations.
This paper focuses on the low frequency shot noise which arises when the pulse excitations are partitioned by a single scatterer in
the framework of the Photo Assisted Shot Noise (PASN) theory. As a unique tool to count the number of excitations carried per pulse,
shot noise reveals that pulses of arbitrary shape and arbitrary charge show a marked minimum when the charge is integer.
Shot noise spectroscopy is also considered to perform energy-domain characterization of the charge pulses. In particular it
reveals the striking asymmetrical spectrum of Lorentzian pulses. Finally, time-domain information is obtained from Hong Ou Mandel like
noise correlations when two trains of pulses generated on opposite contacts collide on the scatterer. As a function of the
time delay between pulse trains, the noise is shown to measure the electron wavepacket autocorrelation function for integer
Lorentzian thanks to electron antibunching. In order to make contact with
recent experiments all the calculations are made at zero and finite temperature.

\end{abstract}

 \maketitle
This paper addresses the noiseless injection of a small finite number of electrons in a quantum conductor. Indeed,
quantum effects become more and more accessible when only few degrees of freedom are controlled. During the
last thirty years, research in this direction has lead to the possibility to manipulate quantum states with
several degrees of freedom and to entangle particles making possible simple quantum information processing.
Up to now most advances have been obtained in quantum optics with the manipulation of single photons emitted by atoms
or semiconductor quantum dots, and in atomic physics with optical arrays of trapped cold atoms or ions
More recently the manipulation of quantum states has become available in condensed matter systems using
superconducting circuits
and semiconductor
quantum dots.

A recent approach is the manipulation of single charges injected in a quantum ballistic conductors. Realizations
of time controlled single charge sources have been
reported in \cite{Feve07,Blumenthal07,Bocquillon12,Hermelin11,Mcneill11} and considered theoretically in
\cite{Splettstoesser08,Olkhovskya08,Splettstoesser09,Moskalets11,Hack11,Albert11,Sherkunov12,Jonckheere12} with a particular focus
on the energy resolved single electron source based on a quantum dot \cite{Feve07}. Injecting more than one electron requires a different
practical approach which is the subject of this paper.
We will consider the more general case of coherent trains of few undistinguishable electrons \cite{coherentstates,Keeling06} which opens the way to
entangle several quasi-particles but also to probe the full counting statistics \cite{Lesovik93} with a finite number
of electrons \cite{Hassler08}.
The injection of single or of few electrons injected in a ballistic one dimensional conductor lets envisage to parallel the flying qubit
approach developed for photons \cite{flying_quBits1,flying_quBits2,flying_quBits3}.
As opposed to Bose statistics, the Fermi statistics makes entanglement of several electrons
 injected in a quantum conductor more favorable and the Coulomb interaction, when not screened, the possibility to
 make non-linear interaction more easier than the Kerr effect used for photons. The ballistic conductors can be realized
 using high mobility 2D electrons confined in high quality III-V semiconductor heterojunctions.
 In addition, applying a perpendicular magnetic field gives rise to one dimensional chiral propagation
 along the edge of the samples in the Quantum Hall regime. This, combined with nano-lithographied split gates,
 allows to implement quantum gates similar to
 that used in quantum optics: electronic beam splitters, electronic Fabry-Perot and Mach-Zehnder interferometers \cite{Ji03,Roulleau08}.

But how to inject single to few electrons in a quantum conductor?
Single electron pumps based on the Coulomb blockade of tunneling have been first realized using ordinary metallic systems
\cite{Geerligs92,Pothier92}.
They have potential application in quantum metrology but are not suitable for our purpose. First the conductor is usually not
ballistic. Second the electrons are sequentially injected. The lack of quantum coherence between electron tunneling events makes
the injection of trains of few undistinguishable electrons impossible.
More recently itinerant quantum dots obtained using surface acoustic waves \cite{Shilton96} have been shown able to transport single
to two electrons along a depleted electron channel \cite{Hermelin11,Mcneill11}. These systems may also
have application in metrology.
Because of the long coherence time of the spin, they could be used for spin based qubit operation and to carry
quantum information
between distant dots \cite{Hermelin11,Mcneill11}. Finally, the on-demand injection of a single electron emitted from a single energy
level suddenly risen at a definite energy above the Fermi sea of the leads has
been realized using a quantum mesoscopic capacitor \cite{Gabelli06} in a non-linear regime \cite{Feve07}.
The system can be viewed as the electronic
analog of a frequency resolved single photon source and opens the way for new quantum experiments where single electrons are injected at tunable energy
well above the Fermi energy. However this kind of on-demand emission is not
generalizable to a coherent train of several undistinguishable electrons.

We consider here a powerful and simpler way to inject a coherent train of $n$ electrons in a single quantum channel
of a ballistic conductor. Here spin is disregarded for simplicity.
The principle, proposed by Levitov et al. in \cite{coherentstates,Keeling06} uses voltage pulses
$V(t)$ applied to one contact. For a quantized action
$e\int_{-\infty} ^{\infty} V(t)dt=nh$, $n$ integer and $h$ the Planck constant, exactly $n$ charges are injected.
As the current $I(t)$ at the injecting contact is at all time equal to $I(t)=\frac{e^{2}}{h}V(t)$ this is equivalent
to write $\int_{-\infty} ^{\infty} I(t)dt=ne$.
This is the simplest injection of charge that we can imagine. On the experimental level, it does not require the implementation of
a quantum dot nor the combination of tunnel barriers or of several Quantum Point Contacts. For quantum bit applications this will
improve the reliability as the delicate circuitry simplifies and reduces only to that necessary for quantum logic gate implementation.
On the conceptual level, this apparently naive
approach involves a non trivial Physics related to the deep properties of the Fermi sea. Indeed, injecting $n$ charges
in general does not mean injecting $n$ electrons alone but also involves a collective excitation of the whole Fermi sea: a neutral
cloud of electrons and holes. However
the beautiful theoretical observation of ref.\cite{coherentstates} is that a voltage pulse with Lorentzian shape carrying one electron
is a minimal excitation state free of these neutral electron-hole excitations. More generally a combination of $n$ Lorentzian
having arbitrary shape and position in time but all carrying a unit charge of same sign is a minimal excitation state.
The remarkable result of ref. \cite{coherentstates,Keeling06} has triggered several
relevant theoretical contributions \cite{Hassler07,Hassler08,Vanevic08,Lebedev11,Haack10,Moskalets11}
in which the property and potential use of the Lorentzian voltage pulses are discussed. An experimental implementation has been recently done
\cite{Dubois12}.


\vspace{4mm}
In the following, we explore some properties of the coherent injection of several indistinguishable electrons by periodic voltage pulses
applied on an Ohmic contact and we consider their characterization in energy and time domain using shot noise.
We compare different pulse shapes and also the departure from integer charge values and we include finite temperature effects with the aim to
provide direct comparison to experiments \cite{Dubois12}. The goal being to produce injection with minimal excitation, we will only consider
the injection of charges with same sign as alternate sign charge injection was found in general not suitable (see ref. \cite{coherentstates}).
 Considering periodic pulses allows to use the powerful
Floquet scattering theory approach of Ref. \cite{Moskalets02}.
While Lorentzian shape pulses with n charges per period correspond to exactly n excitations \cite{coherentstates,Keeling06},
pulses of arbitrary shape contain more excitations \cite{Vanevic08} whose number is calculated for sine, square and rectangular wave shape.
We also address the case of non-integer charge pulses which was shown in ref. \cite{Lee93,coherentstates} to be the dynamical analog of
the Anderson orthogonality catastrophe problem \cite{Anderson67}. Here the periodicity introduces of cut-off in the
long time log-divergent increase of electron and hole excitations at long time. The Anderson physics reflects in a marked minimum of neutral excitations for integer charge per period as theoretically observed in Ref. \cite{Vanevic08}.

\vspace{4mm}
In order to quantify the creation of extra electron and hole excitations, we need
a physical quantity counting them. The right quantity is the low frequency shot noise of the electrical
current already considered for dc transport \cite{Lesovik89,Khlus87,Yurke90,Buttiker90,Martin92,Blanter} where electrons
injected by a dc voltage biased contact are partitioned by a scatterer playing the role of an electronic beam splitter. In a quantum optics language,
 this can be viewed as a Hanbury-Brown Twiss experiment (HBT) with electrons. As remarked
in the pioneering works or refs. \cite{Lee93,coherentstates,Keeling06}, shot noise is also expected when the electron pulse trains
pass through a beam splitter. Indeed all excitations are
partitioned irrespective to their charge, and the shot noise counts the sum of electron and hole excitations while the current counts
their difference. The current noise is
then proportional to the number $n$ of injected electrons, plus the number $N_{e}$ and $N_{h}$ of extra electron
and hole excitations created per period, with $N_{e}=N_{h}$ by definition. Using the separation of the voltage pulse
into its mean value, or d.c. part $<V(t)>=V_{dc}$ and its ac component $V_{ac}(t)=V(t)-V_{dc}$ as in ref. \cite{Vanevic08},
the physics of voltage pulses
can be conveniently discussed in the framework of Photo-Assisted Shot Noise (PASN) where both theoretical
\cite{Lesovik94,Pedersen98,Rychkov05} and experimental results \cite{Schoelkopf98,Reydellet03} are already available.
At zero temperature, the excess noise $\Delta S_{I}$ characterizing the extra excitations is then the difference between the
PASN noise $S_{I}^{PASN}$ due to $V(t)=V_{ac}+V_{dc}$ and the (transport) shot noise (TSN) $S_{I}^{TSN}$
that we would have with only the dc voltage $V(t)=V_{dc}$
\cite{Lesovik89,Khlus87,Yurke90,Buttiker90,Martin92,Blanter,Reznikov95,Kumar96}.
 As $\Delta S_{I}\propto (N_{e}+N_{h})$ and the current is $I=e\nu (N_{e}-N_{h}+n)=n$, $\nu$ being the frequency, minimizing
the noise at constant current  implies $N_{h}=N_{e}=0$. In terms of photo-assisted effect we will see below that this implies only
absorption or only emission of photon, depending on the sign of the $n$ charges injected.

\vspace{4mm}
The paper is organized as follows. In section I we introduce the basic Physics of the photo-assisted effects in a quantum
conductor with a contact driven by a periodic voltage source in the Floquet scattering approach. We then consider the PASN
and the competition between PASN and TSN when both a dc bias and an ac voltage are applied. We recall the remarkable result
of a singularity appearing in the derivative of the noise when $eV_{dc}=nh\nu$, i.e. for exactly $n$ electrons in average
injected per period $T=1/\nu$ making direct connection with the problem of periodic voltage pulses carrying integer charges.
 We also give the expression of the photo-assisted current.

 In section II we address the comparison between several types of integer
 charge pulses : the square, the sine, the rectangular and the Lorentzian. Using the PASN results of section I, we calculate the number
 of electron-hole pairs excitations via the excess noise. For comparison with experiments the computation is done for both
 zero and finite temperature. The hierarchy of the charge pulses regarding noise production is compared with the hierarchy
 of the pulses based on photo-current production.

Section III addresses the case of non-integer charge pulses. We show that for all type of pulses, the number of electron-hole
pairs rises for non-integer charges but is always minimal for integer charges (and even zero for integer Lorentzian pulses).
For large non integer charge number, the number of neutral excitations quickly decreases for Lorentzian while it
remains finite or increases for other type of pulses.

Section IV provides an energy domain characterization of the charge pulses using shot noise spectroscopy. We compute the PASN as a
function of an arbitrary dc voltage bias which provides a direct measure of the photo-absorption probabilities \cite{Lesovik94,Schoelkopf98,
Reydellet03,Vanevic08,Grenier11}. In particular the Lorentzian
pulses show an asymmetric PASN versus dc bias characteristic of the absence of hole type excitations while sine and square
waves lead to symmetric PASN.

Finally, section V gives a time domain characterization of the pulses by looking at the shot noise
generated by trains of electrons colliding on the scatterer. In analogy with Hong Ou Mandel (HOM) experiments \cite{Hong87},
a time delay between the pulse trains emitted by opposite contacts controls the overlap of electronic wavefunctions
which reflects in noise suppression. Here, the gauge invariance maps the HOM experimental scheme discussed here
to a simpler Hanbury Brown Twiss (HBT) problem with only one driven contact. For colliding single charge Lorentzian pulses
the noise is directly linked to the wavepacket overlap of the injected charges while for other pulse shapes the cloud of
neutral excitations also contributes to HOM noise.

\vspace{4mm}

\textbf{I Floquet scattering description of periodic voltage pulses applied on a contact}

\begin{figure}
  \includegraphics[width=7cm,keepaspectratio,clip]{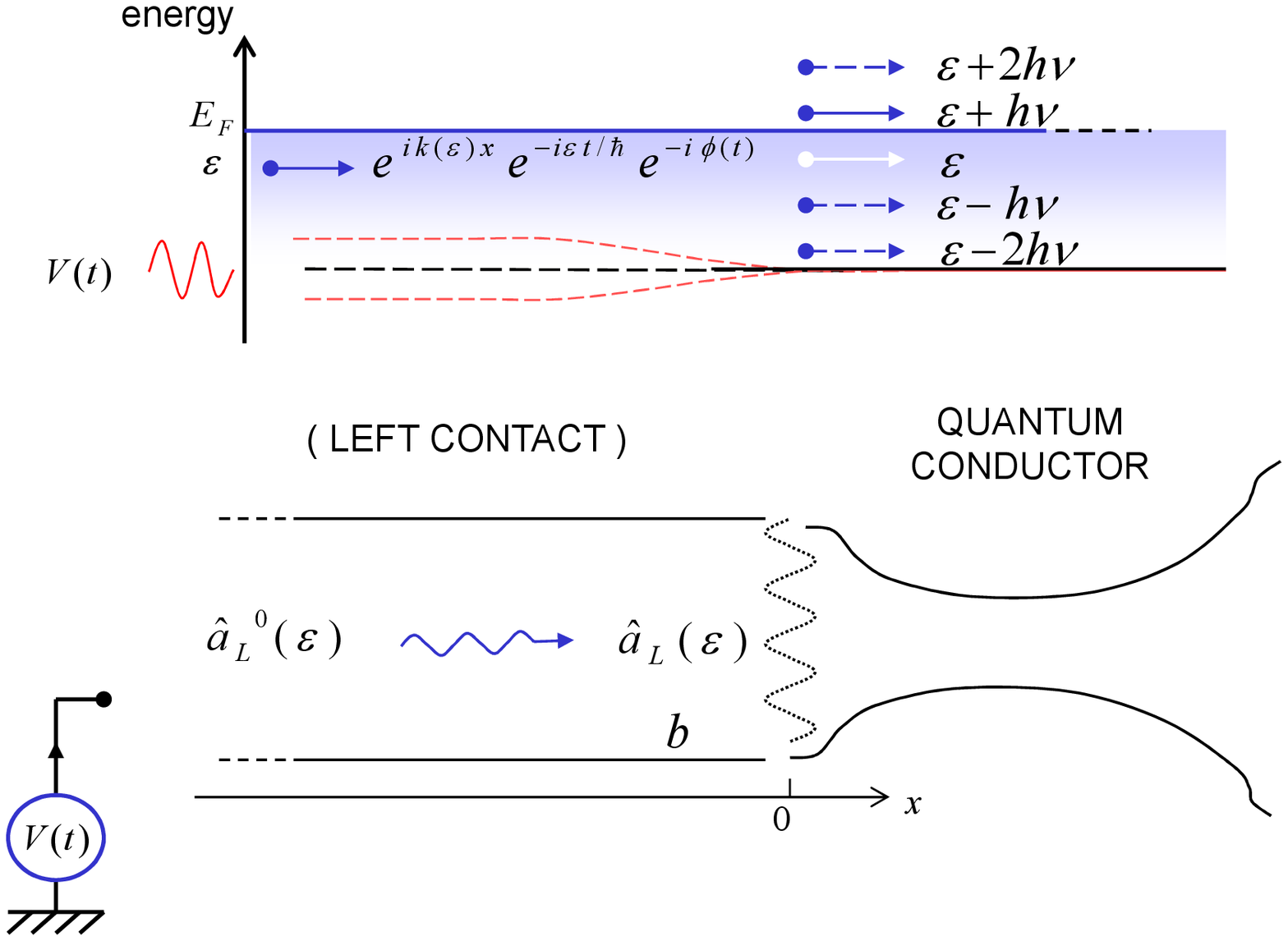}\\
  \caption{Under the effect of an ac voltage, electrons emitted far inside the contact acquire a time dependent phase on their
  way to the scattering region (the quantum conductor). For periodic voltage, frequency $\nu$, the incoming electrons can be described by a
  quantum superposition of states at different energies $\varepsilon + lh\nu$.}\label{scattering}
\end{figure}

We consider a quantum conductor with one contact, say the left (L), periodically driven by a voltage $V_{ac}(t)$ of frequency $\nu=1/T$,
see Fig.\ref{scattering} . Here, without loss of generality we choose the approach of refs.
\cite{Pretre93,Pedersen98} where the periodic potential is assumed to be screened
in all other regions of the quantum conductor. The voltage drop is assumed sharp on the electron wavepacket length but
 smooth on the Fermi wavelength.
The quantum conductor has a small width compared to that of the leads and a small length $L$ compared with the electron
phase coherence length $l_{\varphi}$ such that electrons can propagate coherently over a length $l_{\varphi} \gg L$ on the leads.
The region where electron loose coherence is called the reservoir or contact, following the standard description of the
scattering theory of quantum transport. We assume that the electron coherence time $l_{\varphi}/ v_{F}$ is large compared with
the period $T$. We consider an electron emitted by the left reservoir at energy $\varepsilon$ in a state
 $\sim \exp(ik(\varepsilon)x) \exp(-i\varepsilon t/\hbar)$ with occupation probability $f_{L}(\varepsilon)=1/(\exp(\varepsilon/k_{B}T_{e}) +1)$
 where $T_{e}$ is the electronic temperature and the Fermi energy is the zero energy reference. From the left reservoir to the left
 entrance of the conductor, the electron experiences the potential $V_{ac}(t)$ and acquires an extra term $exp(-i\phi(t))$
 in its amplitude probability with time dependent phase:
 \begin{equation}\label{phi}
    \phi(t)=2\pi \frac{e}{h} \int_{-\infty}^{t} V_{ac}(t^{\prime})dt^{\prime}
\end{equation}
The Fourier transform of:
\begin{equation}\label{pl}
    \exp(-i\phi(t))=\sum_{l=-\infty}^{+\infty} p_{l} \exp(-i 2\pi l \nu t)
\end{equation}
gives the probability amplitude $p_{l}$ for an electron to absorb ($l>0$) or emit ($l<0$) $l$ photons. Eq. (2) expresses that an
electron emitted at energy $\varepsilon$ enters the conductor in a superposition of quantum states at different energies $\varepsilon+lh\nu$. The knowledge
of the $p_{l}$ completely defines the states of the incoming electrons. The magnitude of the $p_{l}$ depends on the reduced quantity
 $\alpha=eV_{ac}/h\nu$ where $V_{ac}$ is the characteristic amplitude of the ac voltage. Combined with the scattering properties of the conductor, all photo-assisted effects resulting from the absorption
 or emission of energy quanta $h\nu$ such has the ac current, the photo-current and the photo-assisted shot noise, etc, can be
 calculated.

The properties of the $p_{l}$ are best expressed in the frame of the Floquet scattering theory \cite{Moskalets02} in which the
continuous energy variable describing states in the reservoir is sliced into energy windows of width $h\nu$, \textit{i.e.}
$\varepsilon \rightarrow \varepsilon +l^{\prime} h\nu$ with $\varepsilon$ now restricted to $[-h\nu,0]$. We can view the
photo-assisted processes as the coherent scattering of electrons between the different energy windows. The scattering matrix
$S(\varepsilon)=\{S_{l l{^{\prime}}}\}$
relates the set of annihilation operators
$\widehat{\textbf{a}}_{L}^{0}(\varepsilon)=\{\widehat{a}_{L}^{0}(\varepsilon +l^{\prime} h\nu)\}$ operating on the states of the left contact
to the set of annihilation operators $\widehat{\textbf{a}}_{L}(\varepsilon)=\{\widehat{a}_{L}(\varepsilon +l h\nu)\}$ of electrons
incoming on the conductor :
\begin{equation}\label{matrix}
    \widehat{\textbf{a}}_{L}(\varepsilon)=S(\varepsilon) \times \widehat{\textbf{a}}_{L}^{0}(\varepsilon)
\end{equation}
with $\{S_{l^{\prime} l} \}=p_{l-l^{\prime}} $.
The unitarity relations $S^{\dagger} S = S S^{\dagger} = \textit{\texttt{I}} $ gives useful relations for the amplitude
probabilities :
\begin{equation}\label{unitarity}
    \sum_{l=-\infty}^{+\infty} p_{l}^{*} p_{l+k} = \delta _{k,0}
\end{equation}

In particular, the sum of the probabilities $P_{l}=|p_{l}|^{2}$ to absorb or emit photons or to do nothing is equal to unity.
As shown below and used in Section IV, the probabilities $P_{l}$ can be inferred from Shot Noise spectroscopy, when in addition to
the ac voltage a tunable dc voltage is applied between the contacts of the conductor. Note that the set of $P_{l}$ does not
contain all the information on the system. Indeed, the products $p_{l}^{*}p_{l+k}$, $k\neq0$, i.e. the non-diagonal part of the
matrix density, enters in the calculation of the
coherence \cite{Grenier11} as discussed in \cite{Grenier12}.

\vspace{5mm}
\textit{Photo-assisted shot noise :}

\begin{figure}
  \includegraphics[width=7cm,keepaspectratio,clip]{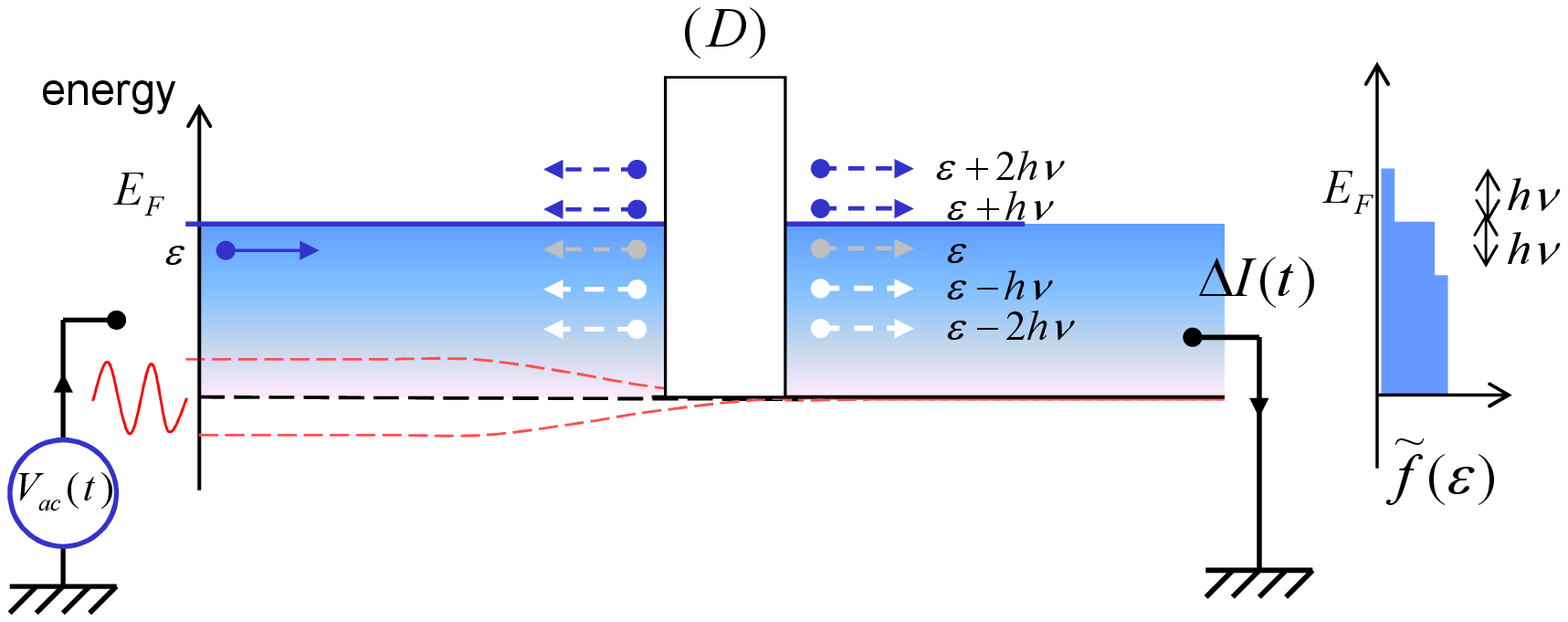}\\
  \caption{Electrons emitted by the left reservoir and pumped by the ac voltage in a superposition of states of different energy are
  scattered. The random partitioning between reflected and transmitted scattering states leads to current noise. The
  noise measures the sum of the number of holes and electrons which are photo-created. On the right is represented the energy
  distribution function $\tilde{f}(\varepsilon)$. It should not be confused with an incoherent distribution function as the
  non-diagonal terms of the density matrix are non-zero.}\label{noisescattering}
\end{figure}

In the following we calculate the photo-assisted shot noise which occurs when the conductor elastically scatters the electrons.
For simplicity we will consider a single mode (or one-dimensional) quantum conductor with transmission probability $D$
as shown in Fig.\ref{noisescattering}.

Noise occurs only when an electron incoming from the left finds no incoming electron from the right or if a hole incoming from the left finds
an incoming electron from the right. Indeed, because of the Pauli principle, or Fermionic anti-bunching, the case where two electrons or two holes
are simultaneously incoming gives no noise. In the former case, the incoming charges are randomly partitioned by the conductor with binomial statistics.
At zero temperature and for zero ac voltage the noise is zero. In presence of an ac voltage electron and hole
excitations are created in the left lead. The photon-created electrons span energies above the right lead Fermi sea and the photon-created holes below the right lead
Fermi sea. All the excitations contribute to partition noise, which is called the Photon-Assisted Shot Noise (PASN). According to ref.\cite{Lesovik94,Pedersen98} the
low frequency current noise spectral density due to photo-assisted process $S_{I}^{PASN}$ is, including the Fermi distributions
$f_{L,R}(\varepsilon)$ of the left and right reservoirs :
 \begin{eqnarray}
    S_{I}^{PASN}=S_{I}^{0} \int \frac{d\varepsilon}{h\nu} \sum_{l=-\infty}^{+\infty} P_{l} \{f_{L}(\varepsilon-l h\nu)(1-f_{R}(\varepsilon))\nonumber \\
     +(1-f_{L}(\varepsilon-l h\nu)) f_{R}(\varepsilon) \} & &
\end{eqnarray}
where $S_{I}^{0}=2 \frac{e^{2}}{h} D(1-D)h\nu$ is the typical scale of the PASN. The complete noise expression is obtained by adding the thermal noise
of the reservoirs: $4 k_{B}T_{e} D^{2} \frac{e^{2}}{h}$. Its origin is the thermal fluctuation of the population in the reservoir and is
not related to partitioning nor to photo-assisted processes.
 In absence of ac voltage, $P_{l}=\delta _{l,0}$, the full noise reduces to thermal noise $4 k_{B}T_{e} D \frac{e^{2}}{h}$ and vanishes at
 zero temperature, as discussed above.

 In order to best extract the Physics, we first consider the zero temperature limit :
 \begin{equation}\label{Si0PASN}
    S_{I}^{PASN}=S_{I}^{0} \sum_{l=-\infty}^{+\infty} |l| P_{l}
\end{equation}
The sum in the right hand side is directly proportional to the number of electrons and holes created respectively above and below the Fermi energy
(chosen as the zero of energy).
To understand this, let us consider for simplicity that only $P_{0}$ and $P_{\pm 1}$ are important and first concentrate on the
absorption process. Electrons below the Fermi surface at energy $\varepsilon - k h\nu$, $k$ positive integer, can be promoted to energy
$\varepsilon - (k-1) h\nu$ with probability $P_{1}$. This global upward shift of the Fermi sea leaves the electron population unchanged
below $E_{F}$ but fills the empty sates of energies $\varepsilon\in [0,h\nu]$ with occupation probability $P_{1}$, above $E_{F}$. Similarly,
in the emission process, electrons are displaced to energies  $\varepsilon - (k+1) h\nu$ with probability $P_{-1}$. The downward shift of the
Fermi sea gives no net change of the population of states with energy $<-h\nu$, while for the energy range $[-h\nu,0]$ the population is now
$1-P_{-1}$. More generally the l-photon processes give electron excitations above the Fermi sea with population $P_{l}$ in the energy range
$[0,lh\nu]$, $l>0$ and holes excitations below the Fermi sea in the energy range $[-lh\nu,0]$, see Fig.\ref{noisescattering}. As the current
of electrons emitted
by the left contact and able to create a $l$-photon electron excitations is $(e/h) l h\nu=l e\nu$ \cite{Reydellet03}, the number of corresponding
electrons created per period is $l P_{l}$. The total number of electron excitations generated per period is thus:
\begin{equation}\label{Ne0}
    N_{e}=\sum_{l=1}^{+\infty} l P_{l}
\end{equation}
and similarly the number of holes :
\begin{equation}\label{Nh0}
    N_{h}=\sum_{l=-\infty}^{-1} (-l) P_{l}
\end{equation}
and, from Eq.(\ref{Si0PASN}), the PASN is :
\begin{equation}\label{Si0}
    S_{I}^{PASN}=S_{I}^{0} (N_{e}+N_{h})
\end{equation}
To conclude this part we must emphasize that the energy distribution function
$\widetilde{f_{L}}(\varepsilon)=\sum_{l=-\infty}^{+\infty} P_{l}f_{L}(\varepsilon - lh\nu)$ depicted in Fig.\ref{noisescattering}
should \textit{not be confused} with an incoherent non-equilibrium population.
If this were the case, even for a perfect lead ($D=1$) one would expect a current
noise associated with the population fluctuation $\propto \int d\varepsilon \widetilde{f_{L}}(\varepsilon)(1-\widetilde{f_{L}}(\varepsilon))$ as for
thermal noise. In the present case the terms $p_{l}^{*}p_{l+k}$ contributing to the non-diagonal part of the density matrix should not
be forgotten. For example they are important for the multiparticle correlations considered by
Moskalets et al. \cite{Moskalets06} or for the concept of electron coherence defined in analogy with quantum optics by Grenier et al.
\cite{Grenier11,Grenier12}. These interference terms contribute to make the low frequency Photo Assisted Shot Noise \textit{strikingly vanishing at unit transmission}. This
was experimentally shown by Reydellet et al. \cite{Reydellet03}. In this work, the theory of quantum partition noise of photon-created electron-hole
pairs, equations (\ref{Si0PASN}-\ref{Si0}), was experimentally checked from weak to large ac excitation and by varying the transmission.
Motivated by this experiment, Rychkov et al \cite{Rychkov05} have theoretically shown that the electron and hole excitations contributing to noise
in Eq.(\ref{Si0}) are not statistically independent. Eq.(\ref{Si0}) was derived by Lee et al. \cite{Lee95}, Levitov et al. \cite{coherentstates} and
later by Keeling et al.\cite{Keeling06}. In the different context of periodic injection of energy resolved single electron and single hole from a quantum dot,
a similar equation has been derived and experimentally tested by Bocquillon et al. \cite{Bocquillon12} for the partitioning of single charges. Finally, as discussed below it is important to note that Eq.(\ref{Si0}) measures the number of electron and hole excitations only accurately at zero temperature.

\vspace{5 mm}
\textit{Photo-assisted and Transport Shot noise :}

  We consider now a dc voltage bias $V_{dc}>0$ added to the periodic ac voltage $V_{ac}(t)$ on the left contact while the right lead Fermi energy is
  kept to zero.
  Let $q=eV_{dc}/h\nu$ be the number of electrons emitted per period due to the dc bias. The total number of
  left electrons participating to noise is $q+\sum_{l=1}^{+\infty} l P_{l}=q+N_{e}$. The number of holes generating partition noise is however reduced
  by the positive shift of the left Fermi energy. For $(n-1)h\nu < q < n h\nu$ the number of holes participating to noise is reduced to
  $\sum_{l=-\infty}^{-n} (-l-q) P_{l}<N_{h}$. The
  shot noise expression (\ref{Si0}) is then changed by replacing the $|l h\nu|$ terms by $|l h\nu+eV_{dc}|$ as originally derived by
  Lesovik et al. and later by Pedersen et al. \cite{Lesovik94,Pedersen98}.
  In absence of ac voltage one recovers
  the transport shot noise $ S_{I}^{TSN}=2 \frac{e^{2}}{h} D(1-D) |eV_{dc}|$.

  The mixed situation with both $V_{ac}$ and $V_{dc}$ leads to
  interesting effects due to the competition between PASN and TSN . They are best displayed using the excess noise where the pure TSN is subtracted
  from the total noise:
 \begin{equation}\label{deltaSi}
    \Delta S_{I}=2 \frac{e^{2}}{h} D(1-D) [\sum_{l=-\infty}^{+\infty} P_{l} |l h\nu+eV_{dc}|-|eV_{dc}|]
\end{equation}
 This would correspond to an experimental situation where the noise measured with $V_{ac}$ on is subtracted from the dc transport shot noise measured
 with $V_{ac}$ off while keeping the dc voltage on \cite{Vanevic08,Dubois12}.

 Finally, we can write reduced units for the excess shot noise $\Delta S_{I}=S_{I}^{0} \Delta N_{eh}$  where :
\begin{equation}\label{Neh}
    \Delta N_{eh}=\sum_{l=-\infty}^{+\infty } |l+q| P_{l} - |q|
\end{equation}
represents at zero temperature the total number $\Delta N_{eh}$ of photon created electrons and holes not contributing to the TSN.

 The derivative of the excess shot noise $\Delta S_{I}$ (or of $\Delta N_{eh}$) with respect to $q$ shows remarkable singularities each time
 $q=eV_{dc}/h\nu$ is an integer $n$.
 At the singularity, the change of slope of the variation of $\Delta S_{I}$ with positive (negative) $V_{dc}$ is
 proportional to $2P_{-n}$ ( $2P_{n}$ ), for $n\neq 0$ and $(2P_{0}-1)$ for $n=0$. Varying $V_{dc}$ thus provides direct information on
 the $P_{l}$. Their direct measure is provided by the second derivative of the noise \cite{Schoelkopf98,KozhevnikovThesis,Shytov05} :
 \begin{equation}\label{secondderivative}
    \partial^{2}\Delta N_{e-h}/\partial q^{2}=(2P_{0}-1)\delta_{q,0} +\sum_{l\neq 0}2P_{l}\delta_{q,-l}
\end{equation}
 Applying a monochromatic sine wave to a contact, the noise singularities at dc voltage multiple of the frequency
 has been observed in a diffusive metallic wire by Schoelkopf et al. \cite{Schoelkopf98} via the second derivative of the noise,
 giving the $P_{l}$ spectroscopy.
Later the controlled suppression by a dc voltage bias of hole or electron excitation contribution to PASN has been discussed and
observed in a Quantum Point Contact by Reydellet et al.
\cite{Reydellet03}. The extracted $P_{l}$ quantitatively agrees with the expected Bessel function.
Inferring the energy distribution of photoexcited electrons from shot noise spectrocopy has been discussed in
 \cite{KozhevnikovThesis} and was compared to tunnel spectroscopy by Shytov \cite{Shytov05}.
In the different context of the energy resolved single electron and hole source
using a mesoscopic capacitor, a similar shot noise spectroscopy was proposed by Moskalets and B\"{u}ttiker \cite{Moskalets11}.
In this same context, the concept of shot noise spectroscopy and equation(\ref{secondderivative}) has been extended to propose a full
tomography of the electron and hole quantum states by Grenier et al. \cite{Grenier11}. The Shot Noise spectroscopy is used in Section IV
to analyze the excitation content of periodic charge voltage pulses.

 Finally, directly relevant to the topics of charge injection, the singularity at $eV_{dc}=n h\nu$ ($q=n$) corresponds
 \textit{exactly to the condition required for injecting $n$ electrons per period}. This is why the presentation of periodic
 charge injection using voltage pulses is particularly relevant and enlightening in the frame of PASN. This singularity is also
 a useful tool to characterize the carrier charge in interacting systems. The superconducting-normal junction where conjugated electron-hole
 Andre'ev pairs carry twice the electron charge have been studied by Torres et al \cite{Torres01}. The finite frequency PASN noise of charge $e$
 and charge $e/3$ partitioned by a QPC was studied in respectively the integer and the fractional quantum Hall regime at $1/3$ Landau level filling
 factor by Cr\'{e}pieux et al \cite{Crepieux04} and later by Chevallier et al. \cite{Chevallier10}.

 For comparison with realistic experimental situation, we provide the excess noise formula at finite electron temperature $T_{e}$. The
 excess noise is $\Delta S_{I}(V_{ac},V_{dc},Te)=S_{I}^{0} \Delta N_{eh}(\alpha,q,\theta_{e})$ where :
 \begin{equation}\label{NehT}
    \Delta N_{eh}(\alpha,q,\theta_{e})=\sum_{l=-\infty}^{+\infty} (l+q) \coth(\frac{l+q}{2\theta_{e}}) P_{l}(\alpha) - q \coth(\frac{q}{2\theta_{e}})
\end{equation}
 and $\theta_{e}=k_{B}T_{e}/h\nu$ is the temperature in frequency units. At finite temperature $\Delta N_{eh}(\alpha,0,\theta_{e})$ no longer
 represents a direct
 measure of $N_{e}+N_{h}$ as the excitations created in the energy range $k_{B}T_{e}$ around the Fermi energy interfere with thermal excitations.
 This reduces the shot noise which therefore misses these excitations \cite{NoteBocquillon}.

\vspace{5 mm}
\textit{Photo-assisted Current :}
Finally, we consider an other photon-assisted effect, the photo-current, which also depends on the probabilities
$P_{l}$. We consider a weakly energy dependent transmission probability $D(\varepsilon)\simeq D+\varepsilon \partial D/\partial \varepsilon$
and neglect the energy dependence of the transmission amplitude phase for simplicity. This situation is known to give a rectification effect
characterized by a quadratic term in the low frequency I-V characteristic of the conductor. In terms of photo-assisted effect this leads to a
dc photo-current whose expression is :
\begin{equation}\label{Iphot}
    I_{ph}=\frac{e^{2}}{h} (h\nu)^{2} \partial D/\partial \varepsilon \sum_{l=-\infty}^{+\infty} l^{2} P_{l}
\end{equation}
One can see that $I_{ph}$ gives information on the $P_{l}$ and can discriminate between different types of ac signal. However as shown in the next sections
it is not as useful as shot noise as it can not distinguish between electron and hole excitations and in particular can not identifies the minimal excitation
Lorentzian pulses. Indeed,
$\sum_{l=-\infty}^{+\infty} l^{2} P_{l}=\frac{1}{T}\int_{0}^{T}|d\varphi(t)/dt|^{2}=(e/\hbar)^2\langle V_{ac}(t)^{2} \rangle$ and $I_{ph}$ gives the same
information than a classical time averaging of $V(t)^{2}$. This quantity is also proportional to the production rate of heat in the contacts
$I(t)V(t)\propto V(t)^{2}$.

\vspace{5 mm}
\textbf{II Integer periodic charge injection}

\begin{figure}
  \includegraphics[width=7cm,keepaspectratio,clip]{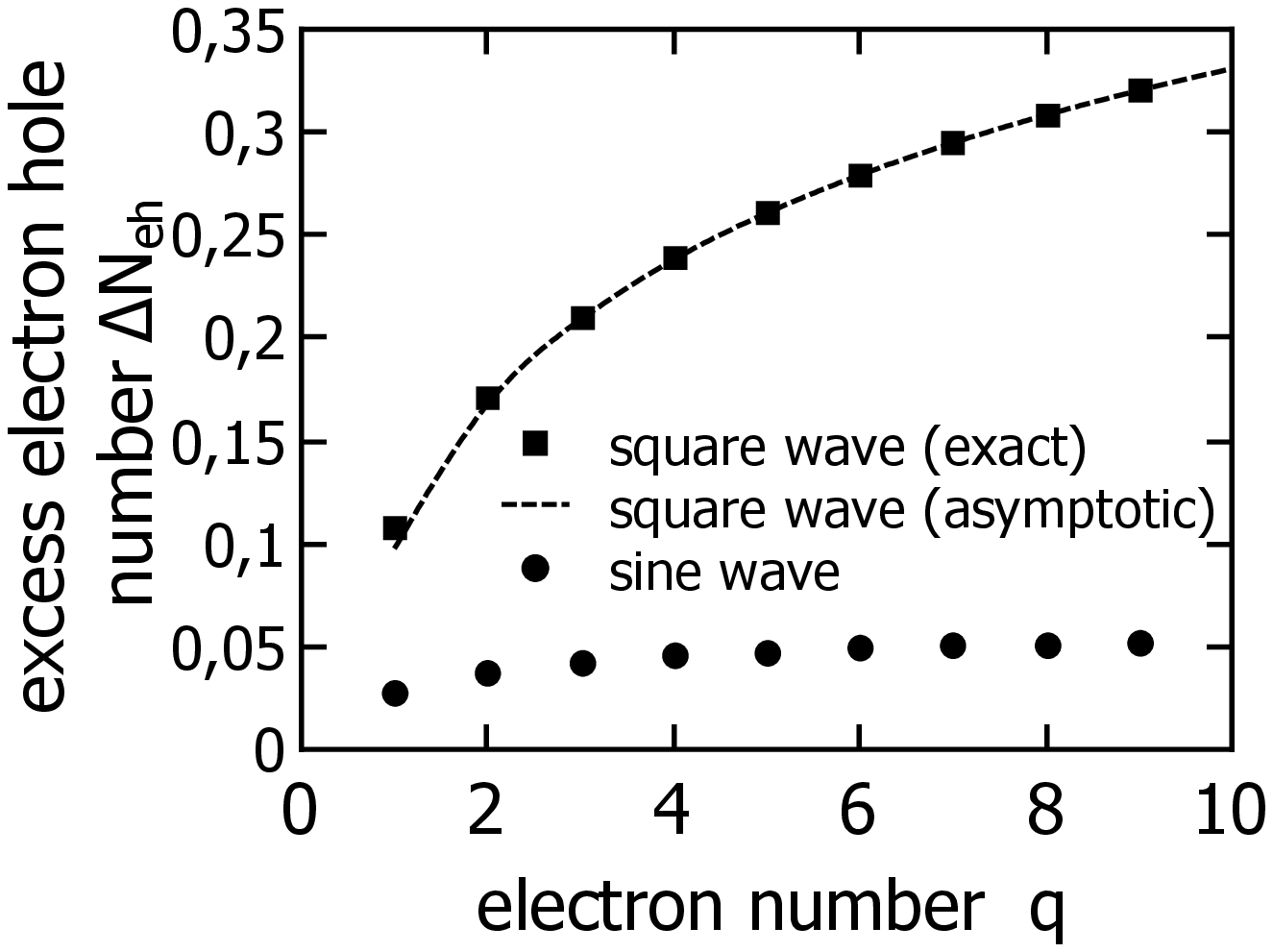}\\
  \caption{Excess electron and hole particle for sine and square wave pulses carrying $q=n$ integer charges per period at zero temperature. The
  asymptotic log divergence of the square for large number, as defined in the main text, is shown as a dashed line }\label{SinSquInt}
\end{figure}
In this part, we give the expression of the probabilities $P_{l}$ associated with various type of pulses carrying
$q = n$ charges per period: the sine, the square and the Lorentzian.
From the shot noise we establish the hierarchy of the pulses in terms of
number of e-h excitations $\Delta N_{eh}$.
We also calculate the photo-current and conclude that
shot noise is the right quantity able to characterize the cleanliness of the charge pulses.
\begin{figure}
  \includegraphics[width=7cm,keepaspectratio,clip]{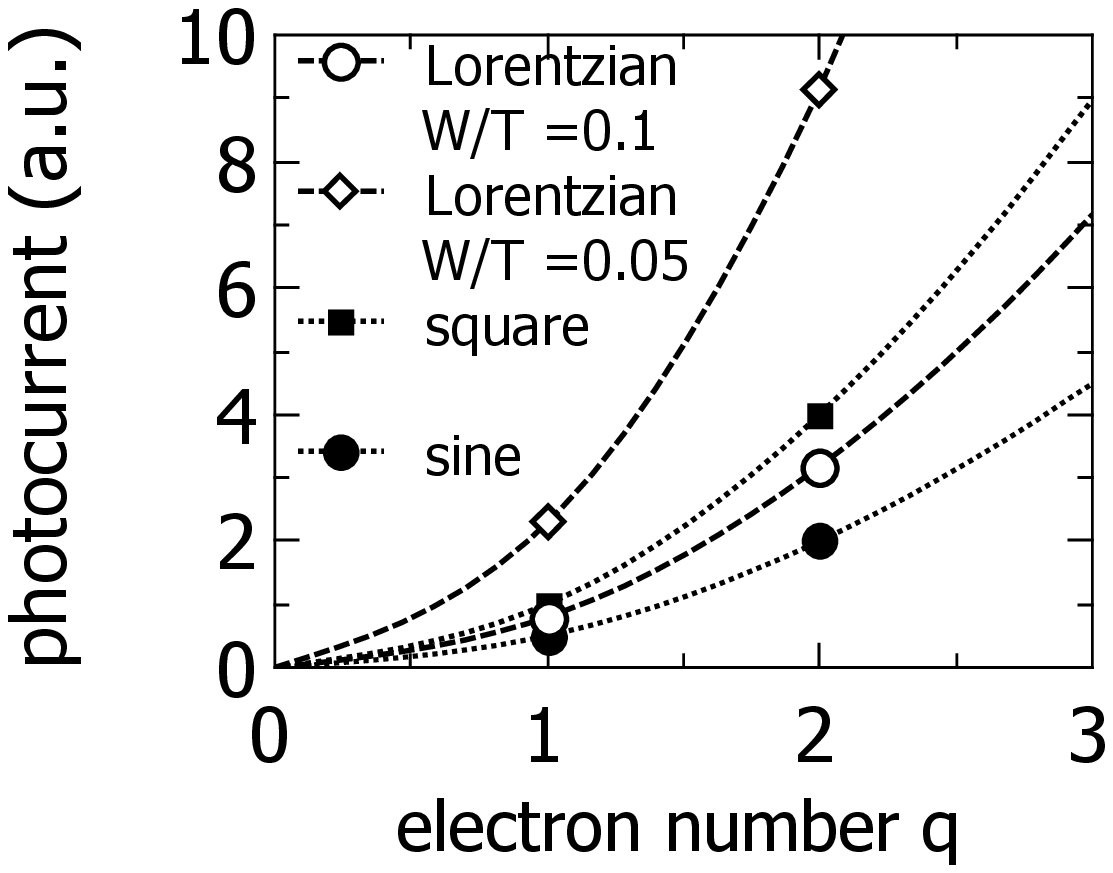}\\
  \caption{Photocurrent in arbitrary unit versus charge per period. We see that this photo-assisted effect can not
  probe the hierarchy of excitation content of different type of pulses. The dashed and doted curves are guide for the eyes.}\label{Photo}
\end{figure}
We consider a periodic excitation carrying a charge $q=ne$ per period $T=1/\nu$. This occurs if $1/T \int_{t}^{t+T} V(t^{\prime})dt^{\prime}= nh\nu$. It
is convenient to decompose the voltage into its mean value $V_{dc} = nh\nu$ and its ac part $V_{ac}$ \cite{Vanevic08,Dubois12}.

For the sine wave $V(t)= V_{ac}(1-cos(2\pi \nu t))$, we have $V_{dc}=V_{ac}=nh\nu/e$. The emission and absorption probabilities are
given by integer Bessel functions with $P_{l}=J_{l}(n)^{2}$ where $P_{l}$ is calculated putting in the time dependent phase only the $V_{ac}$ term
and not the dc voltage part.
Then, we can directly use the PASN shot noise formula with dc bias (\ref{Neh}) to calculate the number
of excitations ($n>0$). Similarly, for the square wave $V(t)=2V_{dc}=2h\nu/e$ for $t\in [0,T/2]$ and $V(t)=0$ for $t\in [T/2,T]$ (mod $T$). The probabilities are given by
$P_{l} = \frac{4}{\pi ^{2}}\frac{n^{2}}{(l^{2}-n^{2})^{2}}$ for odd $l-n$, $P_{l}=0$ for even $l-n$, and $P_{n}(n)$=1/4.
Both the sine and the square wave have symmetric variation around zero voltage which implies symmetric electron and hole excitation creation.
With $P_{l}=P_{-l}$, equation (\ref{Neh}) now writes :
\begin{equation}\label{NehSym}
    \Delta N_{eh}=2 \sum_{l=n+1}^{+\infty} (l-n) P_{l}
\end{equation}
The above expression is useful to provide the asymptotic expression of $\Delta N_{eh}$ at large $n$ for
square wave pulses: $\Delta N_{eh}(n)^{sq}\simeq \frac{1}{\pi^{2}}(\ln(n)+\gamma+2\ln(2)-1)$ where $\gamma$ is the Euler constant.
 For the sine wave case, no log divergence occurs as
$P_{l}\simeq \frac{1}{2\pi l}(e n /2 l )^{l}$ for $l\gg n$.
Figure \ref{SinSquInt} gives $\Delta N_{eh}$ versus the number $n$ of injected electrons for the square and the sine wave.
Clearly the square wave pulse creates more excitations than the sine wave.

We now consider the Lorentzian pulse
whose behaviour is quite different. Strikingly we will see below that the $P_{l}$ associated to the ac part of the voltage pulse are zero for $l<-n$.
The expression for the periodic sum of Lorentzian pulses where each pulse carry $n$ electrons and the Full Width at Half Maximum (FWHM) is $2W$ writes:
\begin{equation}\label{perLor}
    V(t)=\frac{V_{ac}}{\pi}\sum_{k=-\infty}^{+\infty} \frac{1}{1+(t-kT)^{2}/W^{2}}
\end{equation}
where $eV_{ac}=nh\nu$.

The total phase $\Phi(t)$, including the dc voltage part and using reduced time units $u=t/T$ and $\eta = W/T$, gives:
\begin{equation}\label{Phit}
    \exp( -i\Phi(t))=\left(\frac{\sin(\pi (u+i\eta))}{\sin(\pi (u-i\eta))}\right)^{n}
\end{equation}
Remarkably, the above expression has only simple poles in the upper complex plane. This implies that its Fourier transform contains only positive
frequencies. In physical terms, the total voltage only allows electrons to absorb photons, i.e. it induces a global \textit{upward shift of the Fermi sea}
with no hole creation. Going back to the decomposition into dc and ac part of the voltage:
\begin{equation}\label{VacLor}
    eV(t)= n \frac{h\nu}{2} \frac{\sinh(2\pi \eta)}{\sinh(\pi \eta)^{2}+ \sin(\pi u)^{2}}
\end{equation}
with $eV_{dc}=nh\nu$, the amplitude probabilities associated with the only ac part are the Fourier transform of $\exp (-i(\Phi(t) - n\nu t))$ is:
\begin{equation}\label{PlLor}
    p_{l}=\int_{0}^{1} du \left(\frac{\sin(\pi (u+i\eta))}{\sin(\pi (u-i\eta))}\right)^{n} \exp (i(l+ n)u )
\end{equation}
and the probabilities $P_{l}=|p_{l}|^{2}$. From the above consideration on the analyticity of $\Phi$ we see that $P_{(l<-n)}=0$. Note that this is consistent
with vanishing negative frequency Fourier component for the total phase $\Phi$ as the $p_{l}$
differs from them  by of shift of $l$ into $l-n$.

For $n=1$ we find $P_{l}=0$ for $l<-1$, $P_{-1}=\exp(-4\pi \eta)$, and $P_{l}=\exp(-l 4\pi \eta)(1-\exp(-4\pi \eta))^{2}$.
The expression of the $P_{l}$ for integer charge number $>1$ involves the same exponential factor $\exp(-l 4\pi \eta)$ and a
Laguerre polynomial.
The complete expression is given in the next part as a special case of fractional charges $q$ when $q=n$.
Using $\sum_{l=-n}^{\infty} l P_{l}=0$ and equation (\ref{Neh}) gives:
\begin{equation}\label{NehLor1}
    \Delta N_{eh}=0
\end{equation}
The excess shot noise vanishes and only the transport shot noise of $n$ electrons remains. The Lorentzian pulses are therefore \emph{minimal excitation states}
as \emph{no excess electron and hole excitations are created}. The $n$ electronic excitations are however not concentrated on the energy window
$[E_{F},E_{F}+eV_{dc}]$ but they occupy states above the Fermi energy with a weight exponentially decaying on the scale $\sim \hbar/2W$.

We also compare the photo-current for different types of pulses, see Fig.\ref{Photo}.
For the square and the sine wave with $n$ electrons the sum $\sum_{l=-\infty}^{+\infty} l^{2} P_{l}$ is $1$ and $1/2$ respectively. This is consistent with the
hierarchy found using shot noise which showed that the square contains more energetic excitations. For the Lorentzian, the sum is however non zero and given
by $\coth(2\pi \eta) -1$. It strongly increases with the sharpness of the Lorentzian shape.
We conclude that the photocurrent can not characterize the neutral excitation content of charge pulses.
\vspace{5 mm}

\textbf{III Periodic injection of arbitrary charges}

\begin{figure}
  \includegraphics[width=7cm,keepaspectratio,clip]{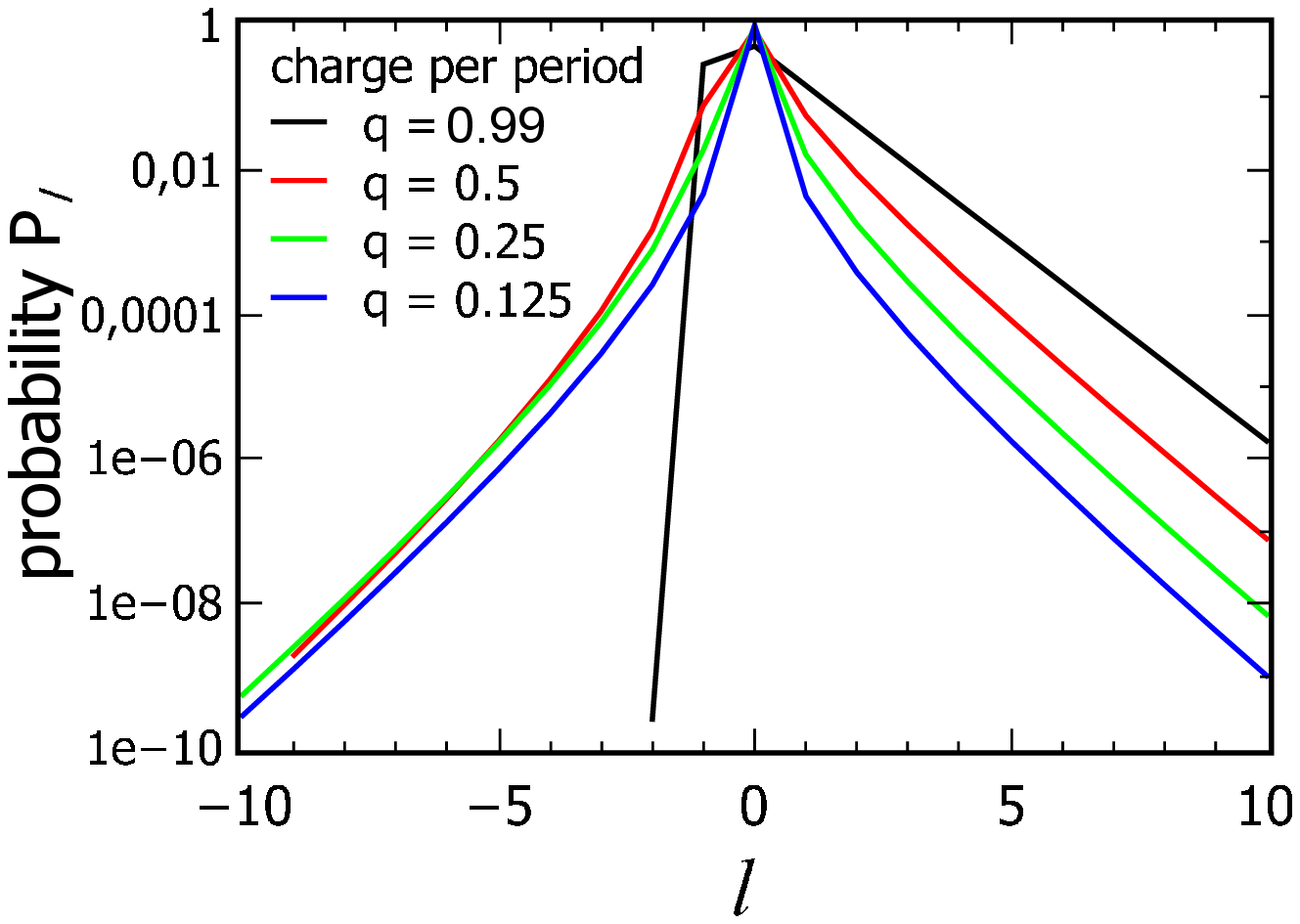}\\
  \caption{absorption $l>0$ and emission $l<0$ probabilities corresponding respectively to electron and hole particle creation for periodic
  Lorentzian pulses of width $W/T = 0.1$ and various charge $q$ per pulse. The asymmetric spectrum for $q=0.99$ reflecting the lack of hole creation quickly
  leads to a symmetric spectrum for $q\ll 1$. Lines connecting the discrete $P_{l}$ values are guide for the eyes.}\label{Plfract}
\end{figure}
In this part, we calculate $\Delta N_{eh}$ for arbitrary charge $q$ carried per period. As in the previous part, we consider the
square, sine and Lorentzian pulse shapes and also include rectangular pulses. We show that $\Delta N_{eh}$ oscillates with $q$ and is locally
minimal for integer $q=n$.
For the sine wave $P_{l}=J_{l}(q)^{2}$ while for the Lorentzian case the calculation of the $P_{l}$ for non integer charges is less trivial.
Physically one may expect that carrying non integer charges
 will involve more complex excitations. Mathematically, we immediately see that the term in the right hand side of equation (\ref{Phit}) is no
 longer analytic
 in the lower plane when $q$, replacing $n$, is not integer. We thus expect a proliferation of hole excitations contrasting with the integer case.
 The calculation is done in the appendix.

Figure \ref{Plfract} shows the $P_{l}$ for periodic Lorentzian pulses of width $W/T=0.1$. The absence of components for $l\leq 1$ when $q=0.99\simeq1$
strikingly contrasts with the case of $q<1$. For small $q$ the $P_{l}$ spectrum is almost symmetrical with $l$ signaling nearly equal electron and hole pair
excitation creation.
\begin{figure}
  \includegraphics[width=6cm,keepaspectratio,clip]{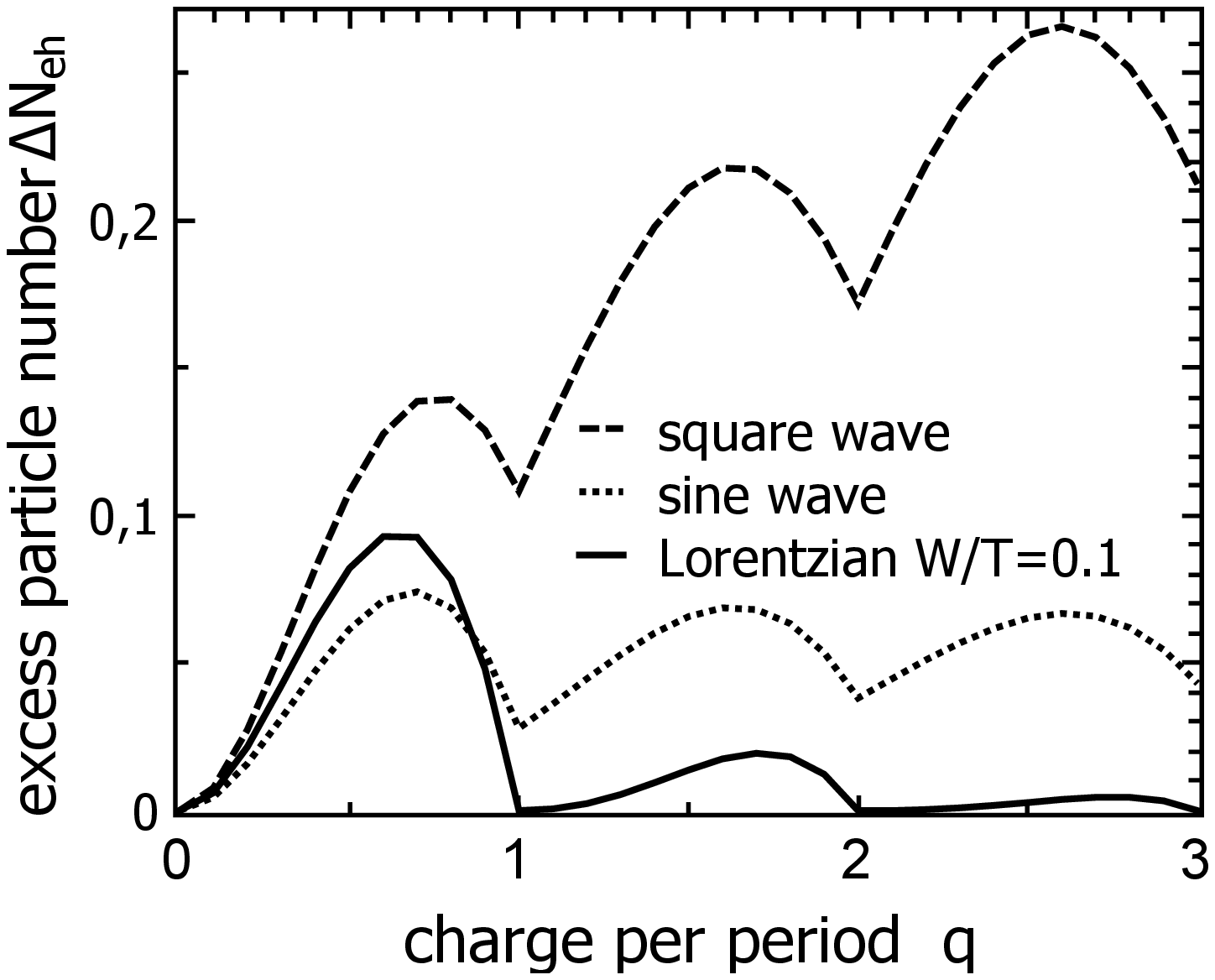}\\
  \caption{Excess electron and hole particle for square, sine and Lorentzian pulses carrying q charges per period. For the Lorentzian the ratio
  $W/T=0.1$}\label{T0lorsinsqu}
\end{figure}
Figure \ref{T0lorsinsqu} shows the evolution of the excess particle number (or excess noise) versus $q$ from $0$ to $3$ for square and sine wave pulses
and for Lorentzian pulses of width $\eta=W/T=0.1$. We observe that, even if the square and the sine do not provide minimal excitations states,
\emph{they do provide a minimum of excitations for integer charges}. This seems to be a remarkable property of the Fermi sea.
A similar figure can be found in Ref. \cite{Vanevic08}.

\begin{figure}
  \includegraphics[width=8.5cm,keepaspectratio,clip]{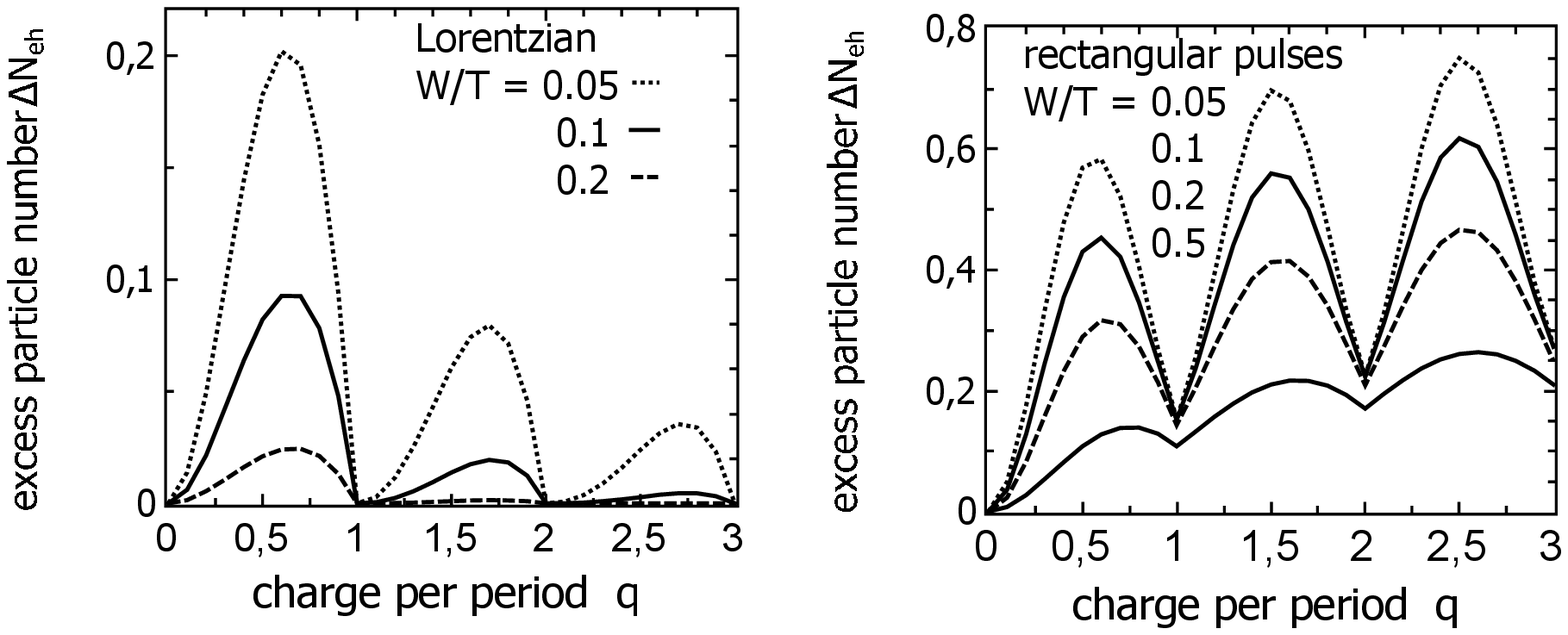}\\
  \caption{Excess electron and hole particle number for Lorentzian (left) and rectangular (right) pulses carrying q charges per period and
  for different width to period ratio $W/T$.}\label{T0lorWs}
\end{figure}

Figure \ref{T0lorWs}, left, shows how the excess particle number evolves for Lorentzian pulses of different width.
We see that for $W/T\geq 0.1$ the particle number becomes exponentially weak. Indeed the potential becomes close to a constant
voltage $V(t)\simeq V_{dc}$. For small width and half integer charge the electron hole excitation number is large but
quickly decreases with q which contrast with the almost constant value found for the sine wave and the increasing value for the square wave.
For comparison, the right graph of figure \ref{T0lorWs} shows the excess particle content of discrete Dirac pulses (or rectangular pulses)
of similar width $W$ (the voltage pulses have amplitude $h/eW$ for the time duration $W$ and zero otherwise). Here again the
excitation content is much larger. It increases with q which contrasts with the Lorentzian pulse behavior which definitely shows the lowest noise.

\vspace{5 mm}

\begin{figure}
  \includegraphics[width=6cm,keepaspectratio,clip]{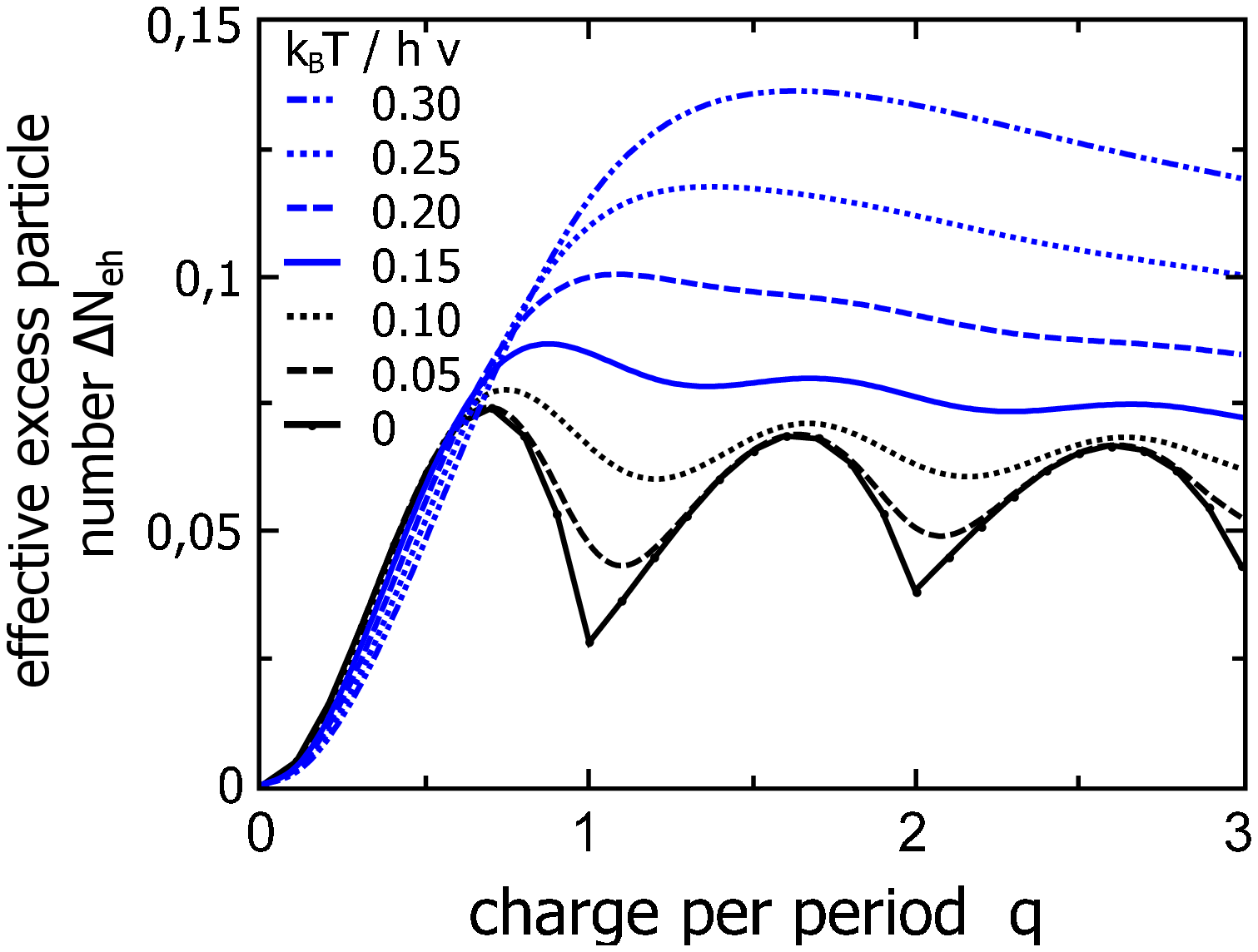}\\
  \caption{Effective excess electron and hole particle for sine pulses carrying q charges per period and for different values of the electron temperature
  $T_{e}$ .}\label{Tsine}
\end{figure}

\begin{figure}
  \includegraphics[width=6cm,keepaspectratio,clip]{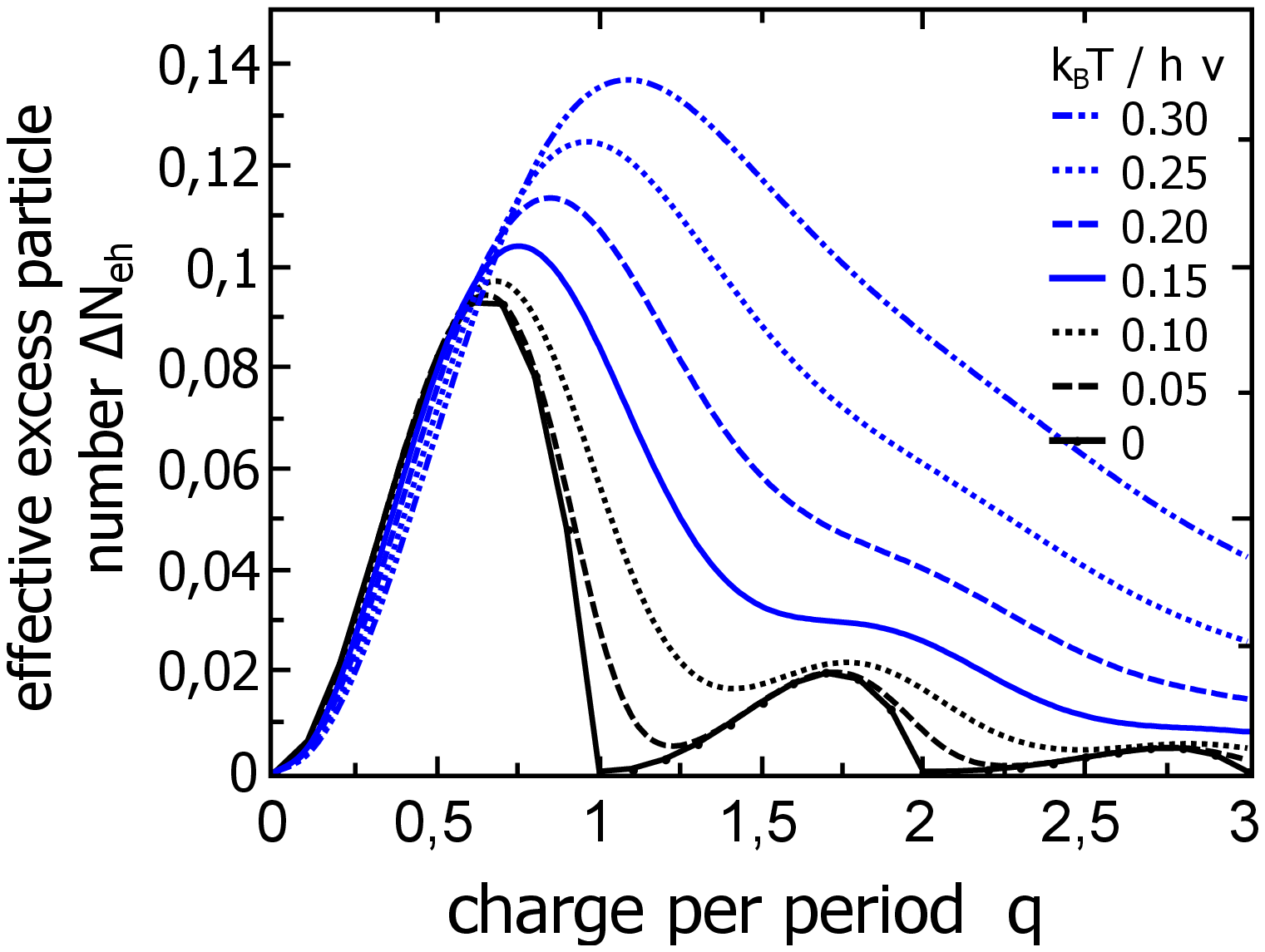}\\
  \caption{Effective excess electron and hole particle for Lorentzian pulses of width $W/T = 0.1$ carrying q charges per period and for different electron temperature $T_{e}$ }\label{Tlor01}
\end{figure}

It is interesting to see how the temperature affects the excess noise. As mentioned previously $\Delta N_{eh}$ no longer measures
the number of excess electron and hole quasiparticles with good fidelity. For simplicity we will keep the same
notation but call it now the effective excess particle number. For sine and square waves there is only one energy scale to compare with the
temperature, i.e. $h\nu$. For the Lorentzian case there are two energy scales $h\nu$ and $\hbar/2W$.
Figure \ref{Tsine} shows the sine wave case. We observe that the minima occur to higher $q$ values. The effect is even more pronounced for the case
of Lorentzian voltage pulses shown in figure \ref{Tlor01} and is probably related to the stronger asymmetry of $\Delta N_{eh}$ with $q$ around $q=n$.
for $\eta=W/T=0.1$.
The oscillations of $\Delta N_{eh}$ are quickly damped by the temperature and when $k_{B}T_{e}>0.2 h\nu$
are almost unobservable.

To end this section, it is worth to note that the non integer charge considered here are injected in a non-interacting Fermionic system (more precisely
a good Fermi liquid where interaction give rise to Landau quasiparticles). Extension to fractional charges in Luttinger liquids has been considered by \cite{Keeling06}. The $e/3$ fractional charge of  fractional quantum Hall edge states has been considered by Jonckheere et al. \cite{Jonckheere05}.

\vspace{5 mm}

\textbf{IV Energy domain: spectroscopy of the e-h pairs excitations}

In this part we use the Shot Noise spectroscopy tool discussed in I to analyze the periodic charged states in the energy domain.
To do this, we compute $\Delta N_{e-h}$ as a function of $V_{dc}$. Here the dc bias $q=eV_{dc}/h\nu$ is no longer linked to the ac amplitude
parameter $\alpha=eV_{ac}/h\nu$. As explained in I, this allows to make a spectroscopy of the electron and hole excitations and infer the
$P_{l}$ from the excess noise variation with $V_{dc}$ or from the first or second noise derivative with respect to $V_{dc}$.
The calculations and graphs are done and displayed at zero and finite temperature.

\begin{figure}
  \includegraphics[width=7cm,keepaspectratio,clip]{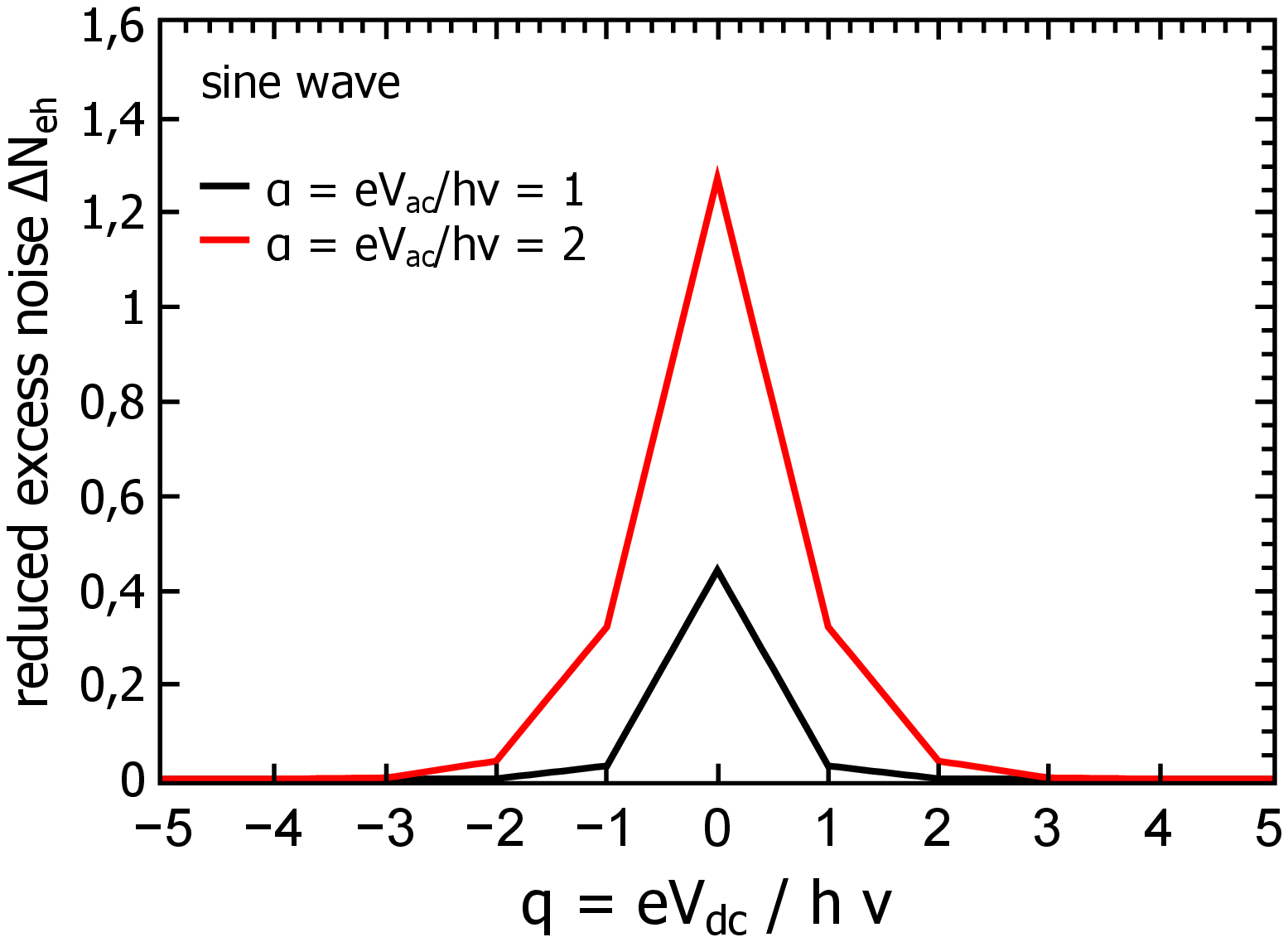}\\
  \caption{Zero temperature excess partition noise versus dc voltage in reduced units for sine amplitudes
  $\alpha=1$ and $\alpha=2$ corresponding to single and double charge pulses when respectively $q=1,2$. }\label{T0sina1a2}
\end{figure}

\begin{figure}
  \includegraphics[width=7cm,keepaspectratio,clip]{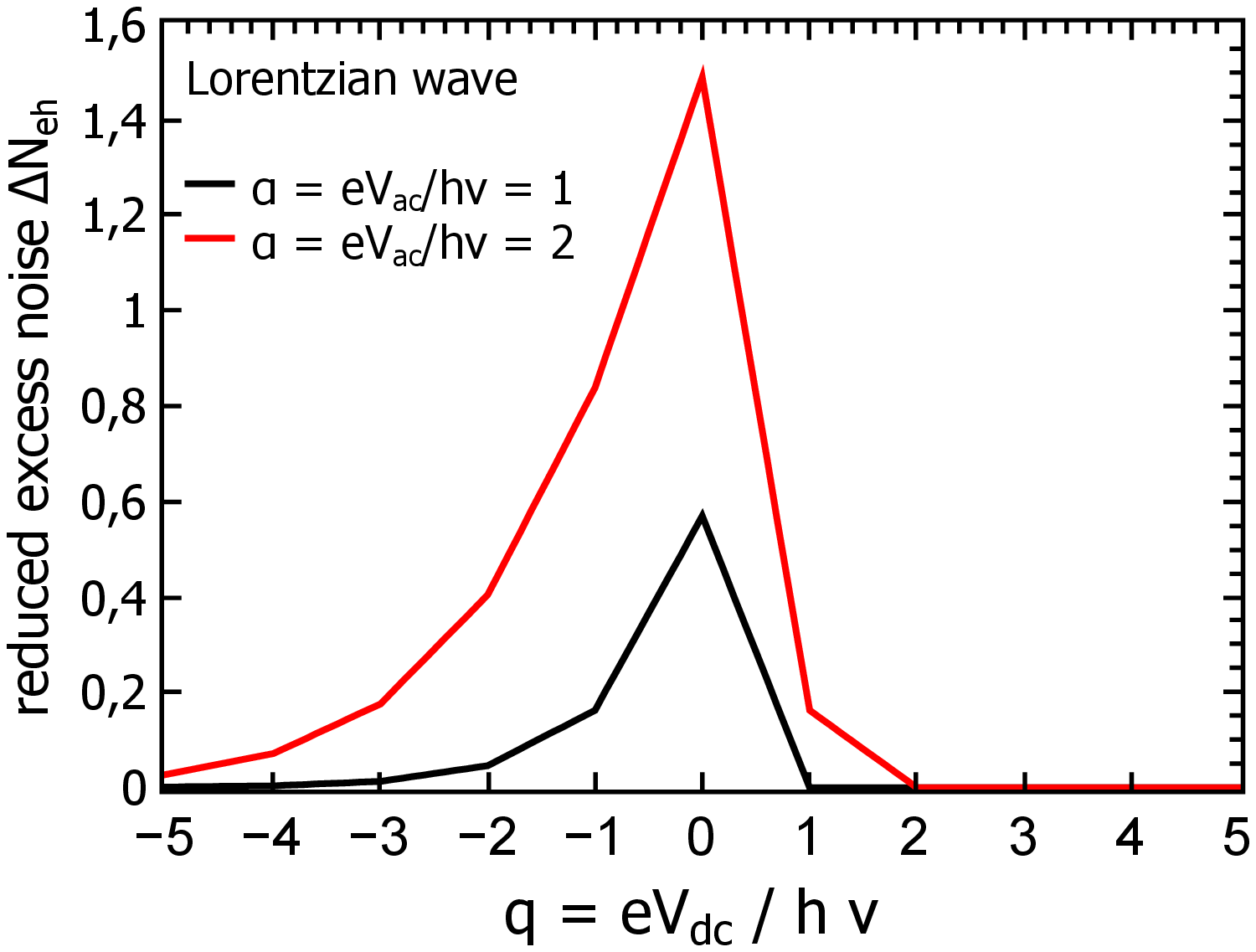}\\
  \caption{
 Zero temperature excess partition noise versus dc voltage in reduced units for
 Lorentzian amplitudes $\alpha=1$ and $\alpha=2$ corresponding to single and double charge pulses when
 respectively $q=1,2$.}\label{T0lora1a2}
\end{figure}

Figures \ref{T0sina1a2} and \ref{T0lora1a2} show the variations of $\Delta N_{e-h}$ versus $q$ at zero temperature for respectively a sine wave and a
Lorentzian wave. For each case we show the curves for two different amplitudes $V_{ac}$ corresponding to $\alpha=1$ and $2$. These amplitudes would
correspond respectively to single and double charge voltage pulses when $q=\alpha=1$ or $2$ respectively.
For sine and square
waves $\Delta N_{eh}(\alpha,q)$
is symmetric with $q$ ( or $V_{dc}$ )  as $P_{l}=P_{-l}$. However it is asymmetric for the case of Lorentzian pulses. Such asymmetry is expected for
pulses whose ac part of the voltage is not symmetrical with respect to zero voltage. But more relevant and striking, for the Lorentzian case
the excess noise is zero for $q>\alpha$ (or $eV_{dc}>nh\nu$), a direct consequence of zero $P_{l}$ for $l<-n$.

\begin{figure}
  \includegraphics[width=7cm,keepaspectratio,clip]{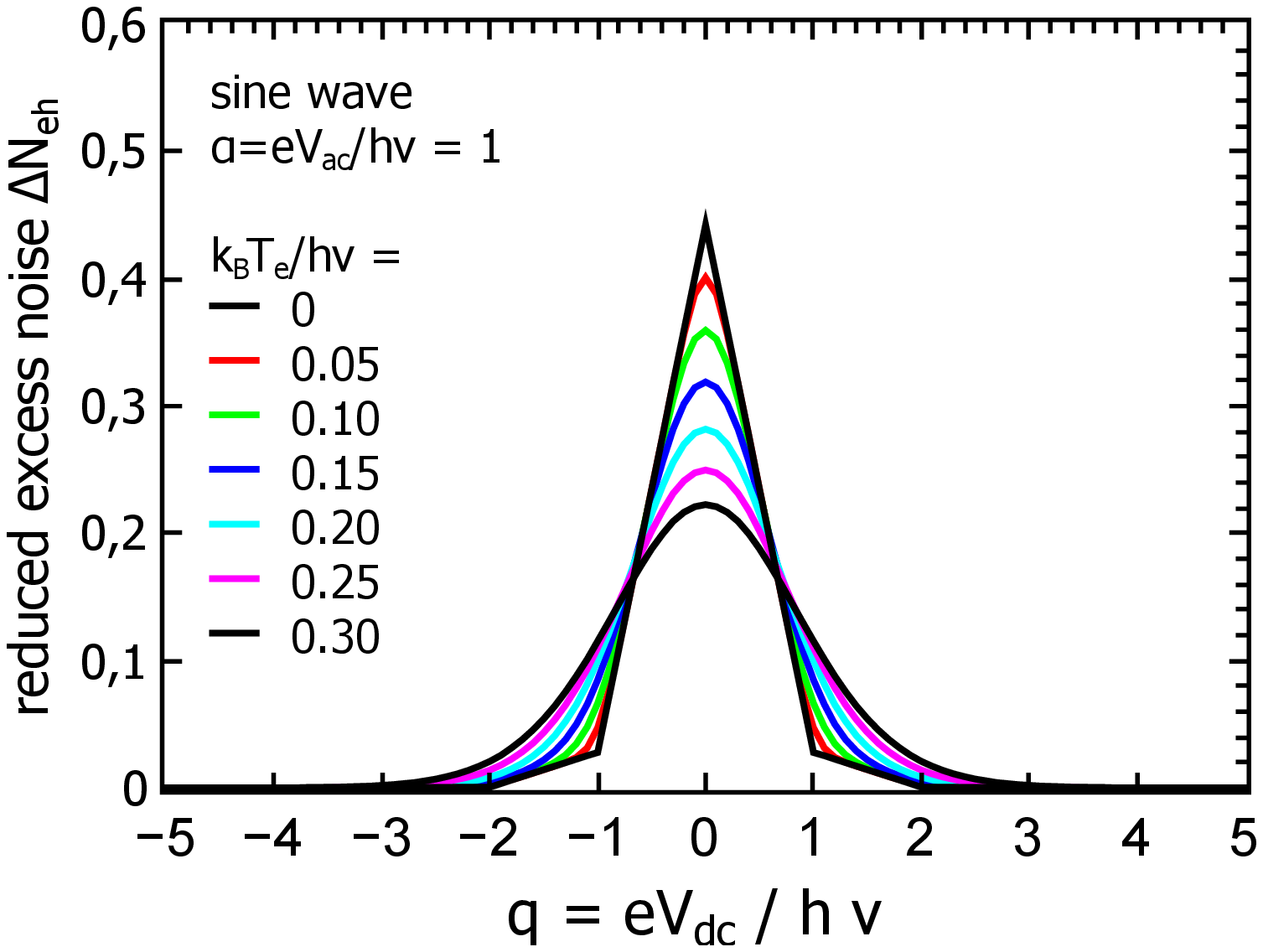}\\
  \caption{Finite temperature excess electron and hole particle noise versus dc voltage in reduced units for sine wave of amplitudes $\alpha=1$. }\label{Tsina1}
\end{figure}

\begin{figure}
  \includegraphics[width=7cm,keepaspectratio,clip]{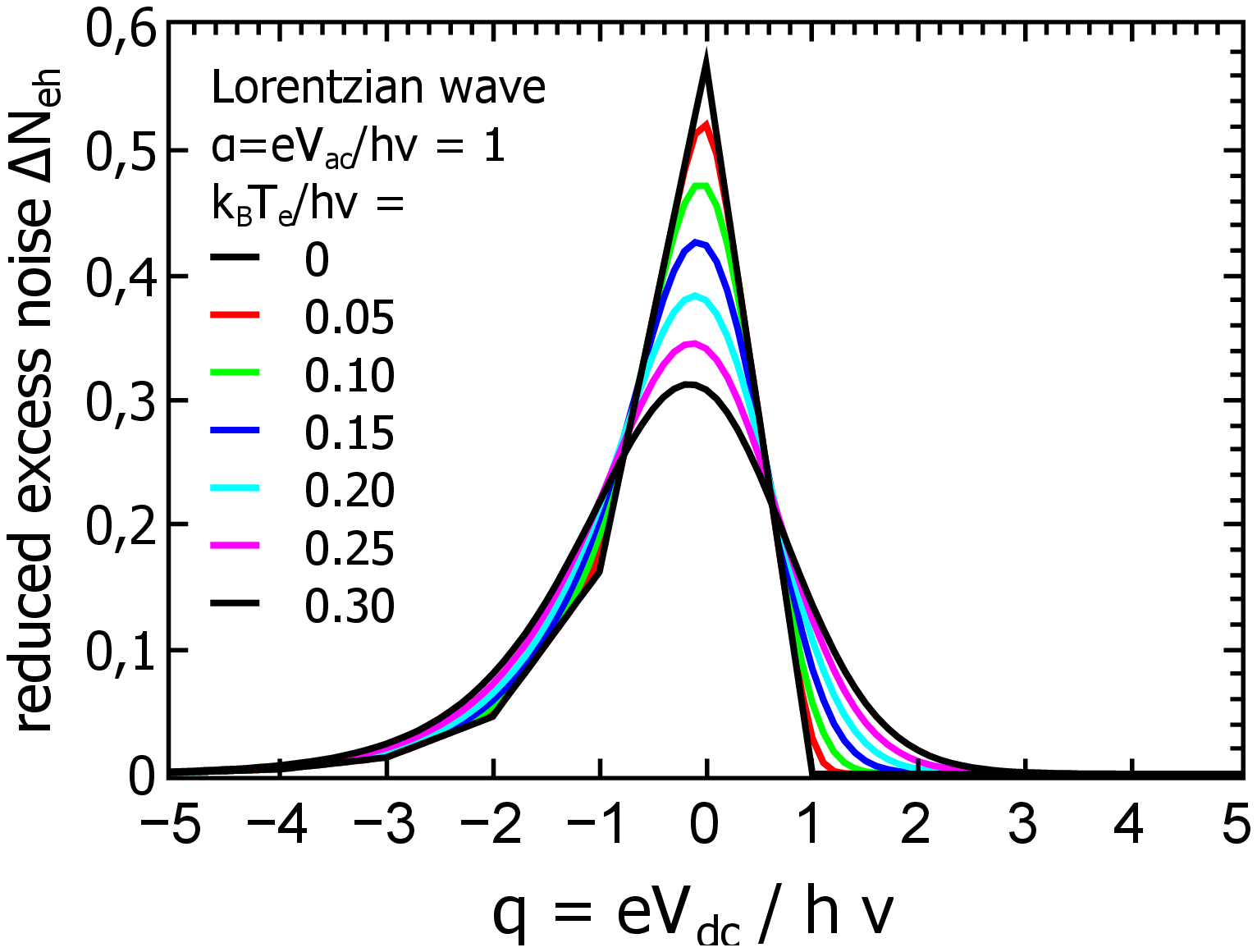}\\
  \caption{Finite temperature excess electron and hole particle noise versus dc voltage in reduced units for Lorentzian waves of amplitude $\alpha=1$. }\label{Tlora1}
\end{figure}

\begin{figure}
  \includegraphics[width=7cm,keepaspectratio,clip]{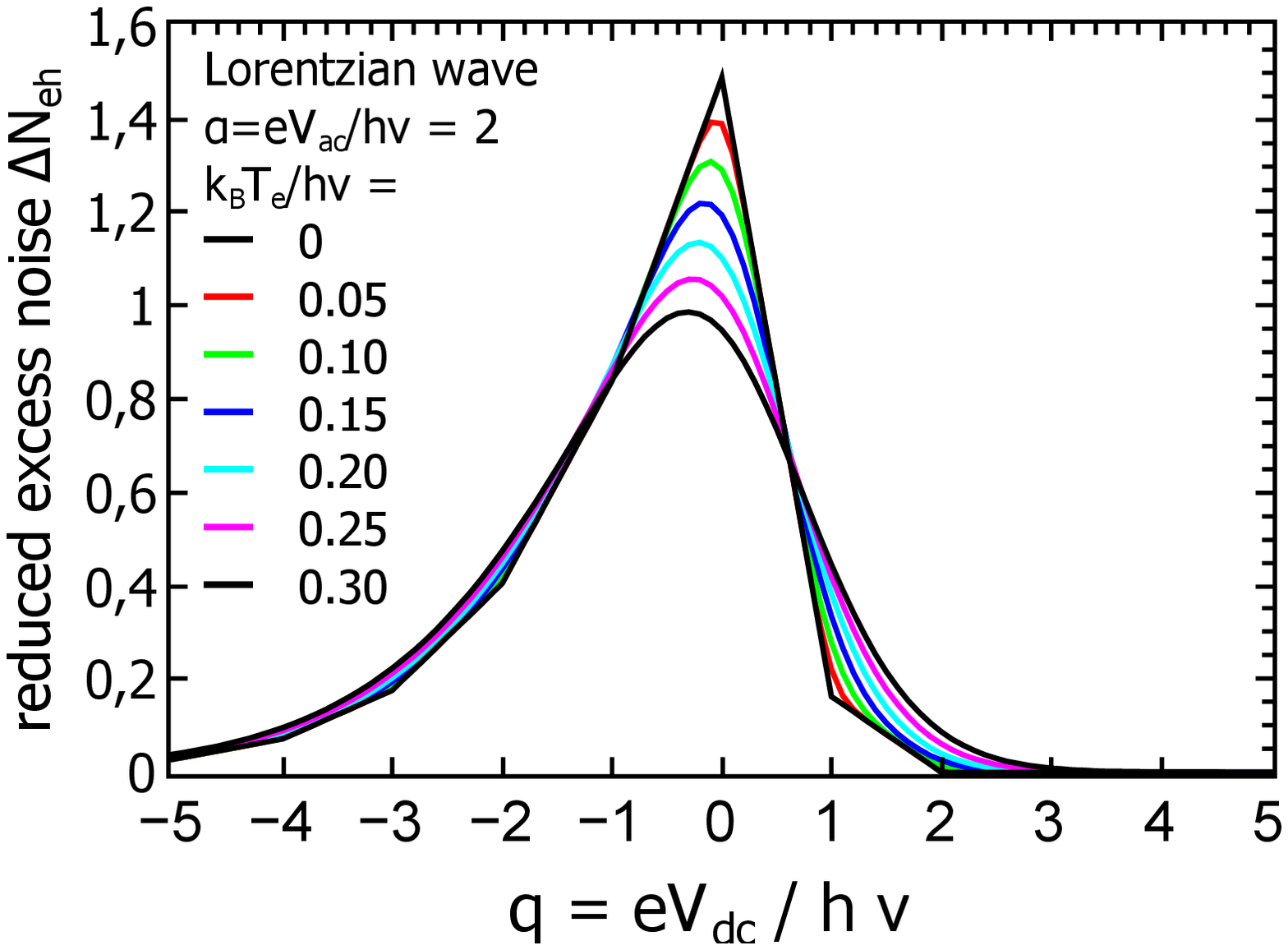}\\
  \caption{Finite temperature excess electron and hole particle noise versus dc voltage in reduced units for Lorentzian waves of amplitude $\alpha=2$. }\label{Tlora2}
\end{figure}

The effect of finite temperature is shown on Figure \ref{Tsina1} for a sine wave of amplitude $\alpha = 1$. Finite temperature calculations
are also shown in figure \ref{Tlora1} and  figure \ref{Tlora2} for Lorentzian pulses of width $W/T=0.1$ with respectively $\alpha=1$ and $2$.
The relevant temperature needed to reveal the asymmetry of $\Delta N_{eh}$ with dc bias for a Lorentzian is given by the width, the smaller the width,
the higher the energy $\hbar/W$ at which we find the contributions of the positive $P_{l}$ responsible for the long tail at negative voltages. The
other temperature scale is given by the ratio $k_{B}T/h\nu $ which controls the smoothing of the singularities at integer $q$.

\vspace{5 mm}

\textbf{V Time domain: shot noise characterization using collision of periodic charge pulses}

The time periodicity imposes that only information on a discrete energy spectrum is available and full characterization must be completed
with a dual time domain information within a period. Here again, shot noise is the useful tool to provide this information.
Inspired by the optical Hong Ou Mandel (HOM) correlation experiment, an electronic HOM analog can be built as a useful tool to infer
the time shape of wave-packets where electrons emitted from two contacts with a relative time delay collide on the scatterer.
A clear relation between shot noise and wavepacket overlap can be made for a single charge Lorentzian pulse.
For other pulse shapes, it is expected that the contribution
of the neutral excitation cloud gives an extra contribution to the shot noise.

In a optical experiment, single photons are emitted from two distinct sources in each of the two input channels of a semi-transparent beam-splitter.
Photon detectors are placed on the two respective outputs and a time-delay between them, sizable with the photon wave-packet,
is introduced. For zero delay, Bose statistics implies a constructive two particle interference where the two photons bunch and exit,
at random, in one of the two output channels. The coincidence events are zero and the particle fluctuations (the noise) is doubled with respect
to a Hanbury Brown-Twiss (HBT) experiment where only one photon at a time arrives on the beam-splitter. The latter situation is recovered when the delay
$\tau$ is much longer than the
size of the wave-packets. For intermediate time delays the noise variation is directly related to the overlap of
photon wavepackets \cite{Hong87,Loudon89}. A similar experiment could be done with electrons with an artificial scatterer
in the form of a controllable beam-splitter, i.e. a Quantum Point Contact. For zero time delay,
Fermi statistics leads to a destructive interference for the probability of finding two electrons in the same output channel. In terms
of charges counted by the detector (here the contacts of a conductor) there is always a charge arriving in each contact and
consequently no current fluctuations.
The noise increases from zero at $\tau=0$ to the single particle noise value (similar to the photon HBT case) for time delay larger than the
electron wave-packet. The difference between Fermion and Boson for HOM correlations has been discussed by Loudon \cite{Loudon98}. An electronic HOM
experiment has been proposed by Giovannetti et al. \cite{Giavonnetti08}. HOM shot noise correlation of electron-hole pairs using
two phase shifted ac voltages of weak amplitude ($eV_{ac}<h\nu$) has been theoretically considered by Rychkov et al. \cite{Rychkov05}.
Up to now, the theoretical works have only addressed the case not discussed here of two single charges emitted from two ac driven quantum dot
capacitors. In this case, the HOM noise has been calculated by Ol'khovskaya et al \cite{Olkhovskya08}.
In the adiabatic regime the wavepackets mimic those emitted by a contact driven by Lorentzian voltage pulses carrying a single electron followed by a single hole (note that, except for well separated electron and holes, the alternate charge prevents to have clean excitations \cite{coherentstates}).
More recently, considering the same system Jonckheere et al. \cite{Jonckheere12} have looked to similar HOM correlations and including the effect of electron
and hole collisions. Their analysis used the quantum electron optics frame developed by Grenier et al. \cite{Grenier11,Grenier12}. 
Other recent theoretical works discussed various related situations \cite{Lebedev08,Splettstoesser09,Moskalets11,Hack11}. In particular
ref.\cite{Moskalets11} considered a mixed situation where electrons emitted form a quantum dot capacitor collide with electrons from a
voltage pulse source. Here we consider only voltage pulse electron sources. This regime was also addressed by in C. Grenier's thesis \cite{Grenier11}.

Practically, two generic situations,
different in their geometry provide similar information: the zero magnetic field case for which the reflected path physically coincides
with the input path
(this corresponds in optics to the mirror facing the two sources and detectors); the high magnetic field case, using the edge states
of the Quantum Hall regime where chiral propagation allows to geometrically separate input and output channels. The first case, provides
simpler interpretation as the interaction are well screened (for example the electrons reservoirs formed by a 2DEG is a good Fermi liquid). The second
case better mimics the optical geometry but the Coulomb interaction between co-propagating quantum Hall edge channels leads to
fractionalization of the injected pulses, see \cite{Grenier12}.

In the following, the HOM shot noise of colliding trains of electron pulses is used as a tool to investigate the
  time shape of electron wavepackets. For simplicity we will consider, as above, a two terminal geometry in zero magnetic field. The result can be
  directly applied to the edge state geometry. Indeed, the noise of a voltage pulse electron source vanishes for unit transmission whatever the pulse shape is.
The autocorrelation and cross-correlation are thus pure partition noise and have same amplitude.
We consider the periodic injection of charges from both contacts with voltage $V(t)$ and
$V(t+\tau)$ applied respectively on the left and right contacts. The scatterer has transmission $D$.
In order to compute the Shot Noise, we use the above photon-assisted Floquet scattering description,
with the left and right input operators defined immediately at the left and right entrance of the electronic beam-splitter (or scatterer) :
\begin{equation}\label{matrixL}
    \widehat{\textbf{a}}_{L(R)}(\varepsilon)=S(\varepsilon)^{L(R)} \times \widehat{\textbf{a}}_{L(R)}^{0}(\varepsilon)
\end{equation}
with
\begin{equation}\label{sllLR}
    S_{l^{\prime}l}^{L(R)}=\frac{1}{T}\int_{0}^{T} dt \exp(i\phi _{L(R)}(t))\exp(i2 \pi (l-l^{\prime})\nu t)
\end{equation}
and $\phi_{L}(t)=2\pi \frac{e}{h} \int_{-\infty}^{t} V(t^{\prime})dt^{\prime}$ and
$\phi_{R}(t)=2\pi \frac{e}{h} \int_{-\infty}^{t+\tau} V(t^{\prime})dt^{\prime}$.
  Then the shot noise can be calculated as in section II now using both the left and right amplitude probabilities.

   Interestingly, in the present case of energy independent scattering,
  the gauge invariance property valid at all time, makes the electronic HOM problem \emph{formally equivalent} to the calculation of
  the shot noise with a single contact, say the left, biased with the voltage difference $V(t)-V(t+\tau)$ while keeping the right contact at zero.
  It is then obvious to have a zero HOM shot noise for $\tau=0$, the case of maximal wavepacket overlap or maximal antibunching. However, the fact that the gauge invariance allows to map the problem to a simpler problem \emph{does not prevent the HOM Physics to be the underlying process}. Indeed, we will see below that the HOM noise is directly related to the overlap of the electronic wavefunctions for periodic trains of wavepackets carrying a single electron.
  A HOM realization not reducible to voltage drop differences via a gauge transformation
  would for example require the use of two single electron sources based on quantum dots \cite{Feve07}.

  \begin{figure}
  \includegraphics[width=8.5cm,keepaspectratio,clip]{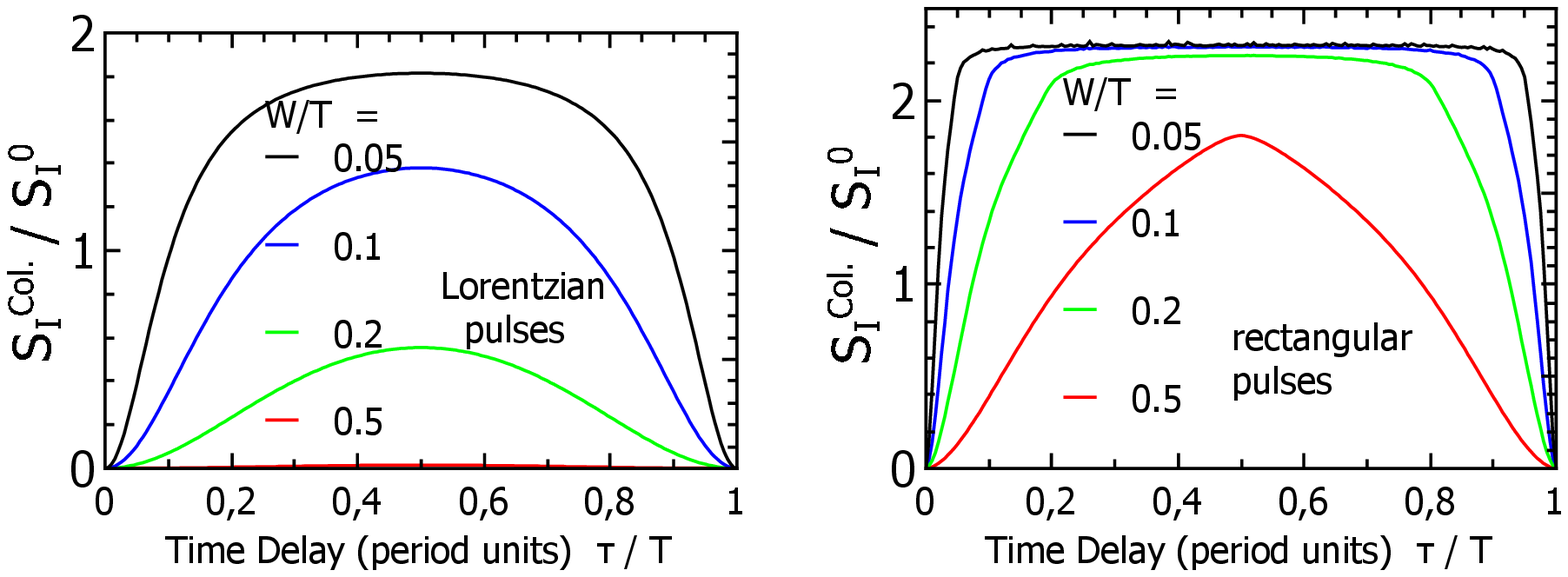}\\
  \caption{Electron shot noise versus the
  time delay $\tau$ for Hong Ou Mandel like correlations for colliding Lorentzian (left) and rectangular (right) pulses  The noise is
  normalized to $S_{I}^{0}$.}\label{HOMLorDir}


\end{figure}

  Figure \ref{HOMLorDir} gives the calculated shot noise in reduced unit $S_{I}^{0}$ versus the time delay $\tau /T$ for the collision of
  single charge pulses per period emitted
  by the left and right reservoirs. The left figure corresponds to Lorentzian pulses of different widths and the right
  figure to discrete Dirac pulses of similar corresponding widths ($V(t)=h/eW$ if $0\leq t \leq W$, $V(t)=0$ for $W<t<T$). For both cases the
  noise starts from zero at $\tau=0$ (perfect antibunching) and is maximum at $\tau=T/2$ where the wavepacket overlap is minimal. For the
  Lorentzian case and $W/T\ll 1$, the wavepackets at $\tau=T/2$ are well separated and the noise is doubled as two electrons (one from the
  left, one from the right) contribute independently to single particle noise at each period. We note however that the noise for
  rectangular pulses exceed the value $2$. Indeed, rectangular pulses (which identify to square waves for $W/T=0.5$) involve a \textit{large
  amount of neutral excitations} which are measured in the HOM noise.

  How far can the HOM noise correlation be used to infer the single charge wavepacket shape in the time domain?
  Ideally, one expects the noise to be given by $2(1-C(\tau))$ where :
  \begin{equation}\label{CTau}
    C(\tau)=| \langle \Psi(t-\tau -x/v_{F})|\Psi(t -x/v_{F}) \rangle |^{2}
\end{equation}
 and $\Psi$ denotes the wavefunction of the excess electron injected right above the Fermi sea by the pulse.
  For Lorentzian pulses, which are clean excitations, a clear positive answer can be given.
  The shot noise of single electron colliding Lorentzian pulse is:
\begin{equation}\label{HOMLorAnal}
    S_{I}^{Coll}(\tau)/S_{I}^{0}= \frac{8 \beta^{2} \sin(\pi \nu \tau)^{2}}{1-2 \beta^{2}\cos(2 \pi \nu \tau)+\beta^{4}}
\end{equation}
The noise is zero for $\tau=0$ and for the case of infinite width $W$ ($\beta=0$). The maximum noise is $S_{I}^{0} \frac{ 2}{\cosh(2\pi W/T)^{2}}$
for $\tau = T/2$.

For $W/T\rightarrow 0$ (or $\beta \rightarrow 0$), it goes to twice $S_{I}^{0}$ as expected for
the partition noise of well separate single charge pulses alternatively emitted
by the left and right contacts. In this limit, we can use the known expression for the wavefunction generated by a single Lorentzian voltage pulse : $\Psi(t)\propto1/(t+iW)$
\cite{Keeling06}. This yields for the
square of the wavefunction overlap $C(\tau) =\frac{1}{1+(\tau / 2W)^{2}}$ and this agrees with the limiting expression of Eq.(\ref{HOMLorAnal}):
 \begin{equation}\label{HOMLorlimit}
    S_{I}^{Coll}(\tau)/S_{I}^{0} \simeq 2(1-\frac{1}{1+(\tau / 2W)^{2}})
\end{equation}
In the appendix we show that Eq.(\ref{HOMLorAnal}) is also proportional to $1-C(\tau)$ even for the case of arbitrary overlap between pulses carrying single electrons. In the small $W/T$ limit,
close expressions have been obtained by Ol'khovskaya et al. \cite{Olkhovskya08} and by Jonckheere et al. \cite{Jonckheere12}
 (see also  C. Grenier Thesis \cite{Grenier11}).
In these works, single charges
whose charge sign periodically alternates are emitted from an ac driven quantum dot capacitor and the
trains of charges collide in a QPC. In the adiabatic ac drive limit, the wavepackets mimic those emitted by Lorentzian voltage pulses on a contact.

Another striking property of the shot noise of colliding single charge Lorentzian pulses is the trivial effect of
the temperature:
$S_{I}^{Coll}(\tau,T_{e})=S_{I}^{Coll}(\tau,0)F(T_{e})$
\begin{equation}\label{HOMLorAnalTempe}
    F(T_{e})=\frac{1-\beta^{2}}{\beta^{2}}\sum_{l=1}^{l=\infty} l \beta^{2l}\coth(\frac{lh\nu}{2k_{B}T_{e}})
\end{equation}
The time delay variation and the temperature variation decouple: \textit{thermal fluctuations do not prevent to get accurate information on the
time shape of the wavepacket } but they only reduce the noise signal. As shown in the appendix this is not true for sine wave pulses and in general
for other pulse shapes.

\begin{figure}
  \includegraphics[width=8.5cm,keepaspectratio,clip]{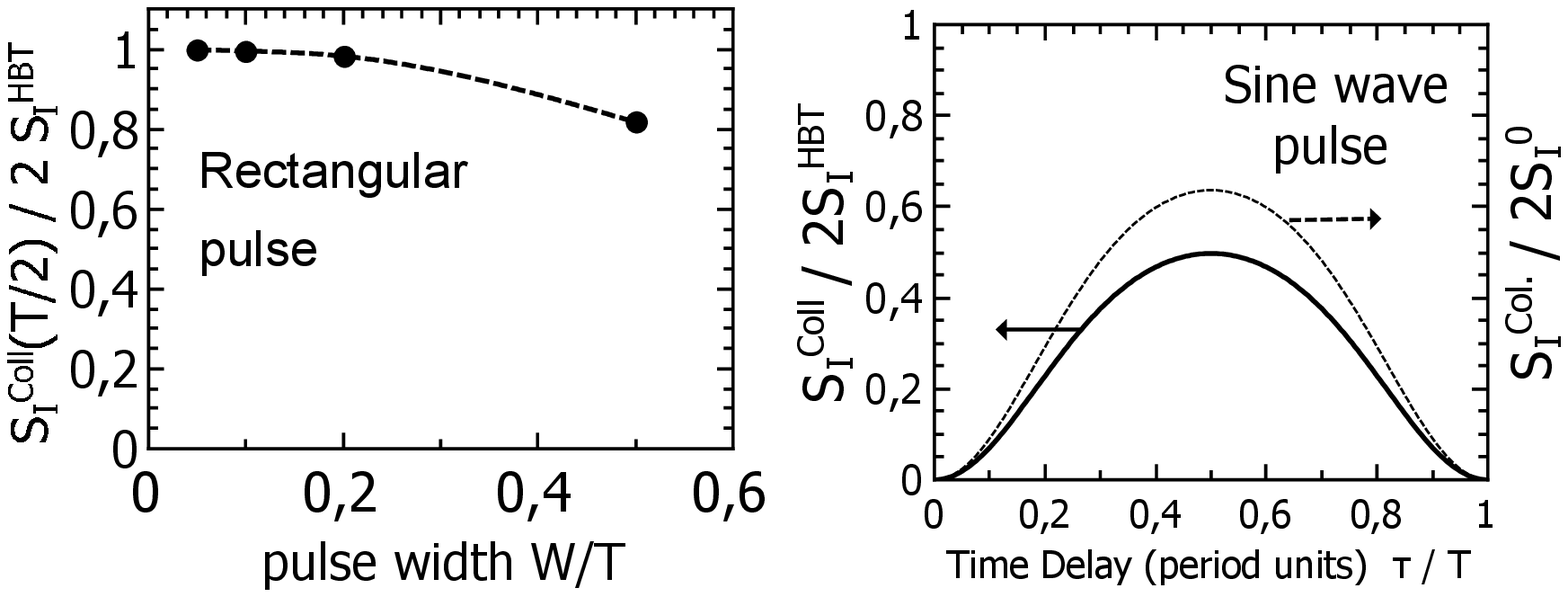}\\
  \caption{left figure: HOM noise at $\tau=T/2$ for a rectangular pulse normalized to twice the partition noise of a single pulse, versus pulse
  width; right figure: HOM noise of sine pulses versus $\tau/T$ normalized to twice the partition noise of single charge (dashed line, right axis) and
  to twice the single pulse partition noise, solid line left axis}\label{FigSineHOM}
\end{figure}

 \begin{figure}
  \includegraphics[width=8.5cm,keepaspectratio,clip]{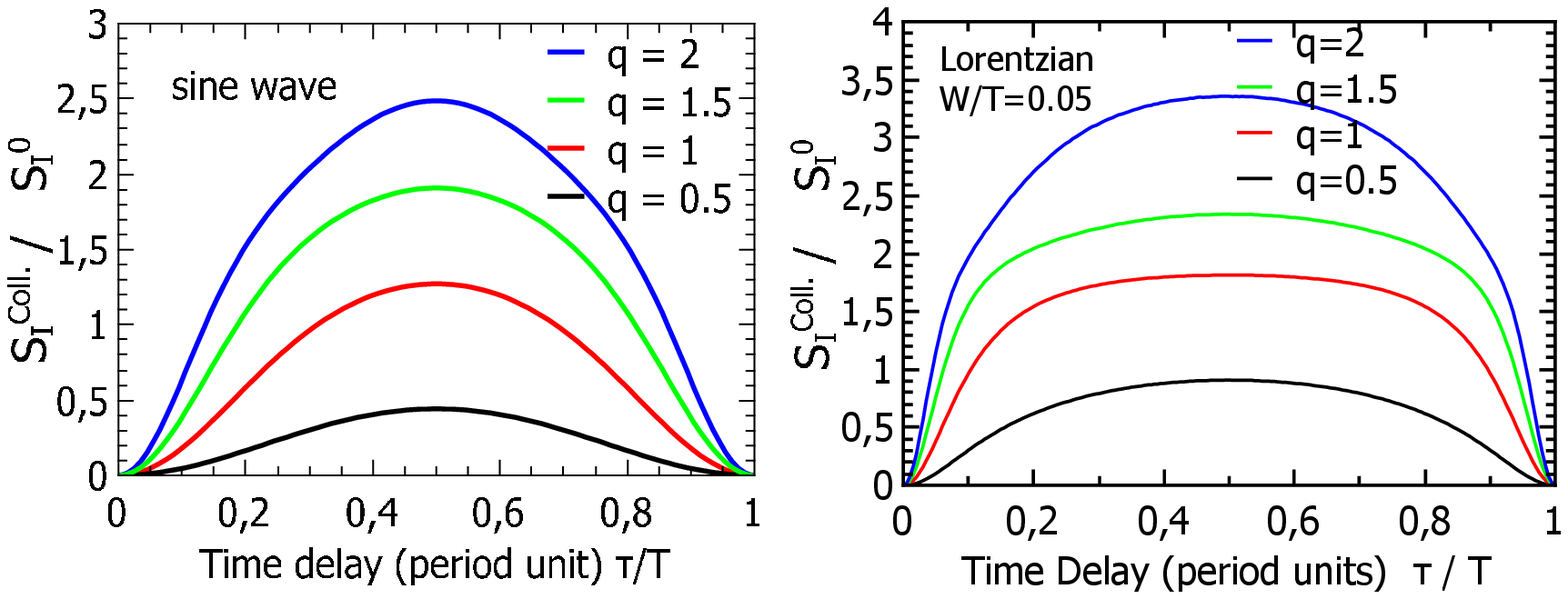}\\
  \caption{HOM shot noise for sine (left graph) and Lorentzian (right graph) pulses versus time delay and
  carrying charges $q=0.5$, $1$, $1.5$ and $2$. }\label{FigSineFracLorHOM}
\end{figure}
All theses remarkable properties demonstrate that, \emph{for integer charge Lorentzian, the HOM Shot Noise does give
information on the time domain shape of the charge wavepacket}. This property is no longer verified for the case of non Lorentzian pulses as the
neutral excitations contributes. The left graph of Figure \ref{FigSineHOM} shows the value of the HOM noise of Dirac pulses time shifted by $T/2$
normalized to the partition noise of single pulses, denoted $S_{I}^{HBT}$. We see that when the overlap between pulses becomes
negligible, the ratio tends to twice the single pulse partition noise. This doubling is expected as now two separate pulses are
partitioned per period. For large overlap, the ratio found in Fig.(\ref{FigSineHOM}) is well smaller than two because of antibunching. An inherent large
overlap is also found for a sinewave, right graph of Fig.(\ref{FigSineHOM}).

Finally we have calculated the HOM noise of Lorentzian pulses for larger charge, including fractional.
For non integer charge pulses we do not expect to get simple information on charge pulses in the time-domain because of the large content of neutral
excitations expected even for the Lorentzian case. The left and right graphs of Figure \ref{FigSineFracLorHOM} show the shot noise for colliding
sine wave and Lorentzian pulses for $q=0.5$, $1$, $1.5$ and $2$. For both pulse shapes the variation
with $\tau$ for $q=2$ suggests a double structure related to the interference of several electrons. A detailed study of colliding pulses for large $q$
will be given elsewhere \cite{Glattli12}.

To conclude this part, we have shown that the HOM electronic shot noise correlation provides meaningful time-domain information on the
wavepacket. This is true however only for single charge Lorentzian which is a clean excitation. Other kind of electron generation is
accompanied by a cloud of neutral excitations which contribute to the HOM shot noise and prevent to get time domain information on the charge
part of the excitations.

\vspace{5 mm}

\textbf{Conclusion}

We have studied in detail the physics of periodic voltage pulses carrying few electrons with the aim to give a useful basis for comparison with experiments
\cite{Dubois12}. Shot noise, as emphasized in the pioneering papers on Lorentzian voltage pulses, is a tool of choice to characterize the charge pulses.
We have exploited the periodicity to show the intimate connection with the well established physics of Photo Assisted Shot Noise using the powerful Floquet scattering approach. We have used the known properties of PASN as a direct measure of the electron and hole excitation number and, when combined with transport
shot noise, as a tool for spectroscopy of the excitations.
In the later case this provides an energy domain characterization of the pulses accessible to experiments.
This information has been supplemented by a time domain study of the charge wavepackets using HOM-like pulse collisions.
The gauge invariance was used to map the charge pulse collision problem to the partitioning of neutral pulses.
The comparison of different pulse shapes emphasizes the peculiar nature of Lorentzian pulses with integer charge
as ideal electron source. We have also considered pulses with arbitrary charge and show that they always contain a large number of
neutral excitations, even for Lorentzian, a manifestation of dynamical orthogonality catastrophe for Fermions.
For large arbitrary $q$, the low noise property of Lorentzian remains remarkable as the
neutral excitation content rapidly decreases with $q$ while it shows a logarithm increases for rectangular shape pulses of same width,
or remains almost constant for sine wave pulses.

Integer charge Lorentzian pulses deliver a noiseless source of few electrons. One expects many applications in quantum physics. For example
by repeating the injection of clean coherent packets with same electron number and measuring the charge arrived in contacts, the full
counting statistics of few electrons partitioned in quantum conductors become accessible. One may also envisage to explore the Fermionic
statistics of multiple electrons colliding of a beam splitter or interfering in Mach-Zehnder interferometers. Integer charge Lorentzian pulses
are also the appropriate source for realizing electronic flying qubits and to entangle few electrons
in ballistic conductors. Finally the powerful approach of Lorentzian pulses could find similar applications in
fermionic cold atom analogs.

Support from the ERC Advanced Grant 228273 MeQuaNo is acknowledged by D.C.G., J.D., T.J. and P.R.. The support from ANR-2010-BLANC-0412 is acknowledged
by C.G. and P.D..

\vspace{5 mm}

\textbf{Appendix}
\vspace{5 mm}

\textit{Lorentzian voltage pulses carrying arbitrary charge:}

In this part we calculate the amplitude probabilities $p_{l}$ for Lorentzian pulses with arbitrary charges. We start with equation
(\ref{PlLor}) with $n$ replaced by $q$ :
\begin{equation}\label{PlLorfrac2}
    p_{l}=\int_{0}^{1} du \left(\frac{\sin(\pi (u+i\eta))}{\sin(\pi (u-i\eta))}\right)^{q} \exp (i2\pi (l- q)u )
\end{equation}
Here, the physical decomposition of the voltage into dc and ac part is
essential to avoid ill mathematical behavior. While the Fourier transform of $\exp( -i\Phi(t))$ leads to divergent logarithmic terms as
 $\Phi(T^{-})-\Phi(0^{+})=2\pi q$ is not a multiple of $2\pi$, that of the phase associated to the ac part is well defined
 as $\phi(T^{-})-\phi(0^{+})=0$ .

Defining $z=\exp(i2\pi u)$ and $\beta=\exp(-2\pi \eta)$

\begin{equation}\label{Plzed}
    p_{l}=\frac{\exp(i2\pi q)}{i2\pi}\oint \frac{dz}{z} z^{l} \left(\frac{1-\beta z}{1-\beta \overline{z}}\right)^{q}
\end{equation}
As $\beta <1$ and the integration contour over $z$ is on the unit circle, both the numerator and the denominator are convergent series
of positive powers of $\beta z$ and $\beta \overline{z}$ respectively. The integral reduces the double series expansion
to a single power series of $\beta$
by imposing that the power of $z$ under the integrand is zero.
After some algebra we get :
\begin{equation}\label{PlLorfracPos}
    p_{l}=q e^{iq\pi} \beta ^{l} \sum_{k=0}^{\infty} (-1)^{k} \frac{(q+l+k-1)!}{(k)! (q-k)! (l+k)!} \beta ^{2 k}
\end{equation}
for $l>0$, and
\begin{equation}\label{PlLorfracNeg}
    p_{l}=(-1)^{l} q e^{iq\pi} \beta ^{|l|} \sum_{p=0}^{\infty} (-1)^{p} \frac{(q+p-1)!}{(p)! (q-|l|-p)! (|l|+p)!} \beta ^{2 p}
\end{equation}
for $l<0$. Interestingly, while for non integer $q$ the sum are infinite (but quickly convergent), for $q$=n the sums reduce to the expected polynomial
expressions of order $n-1$ and the terms for $l<-n$ vanish. Indeed when the terms $(q-k)!$ and $(q-|l|-p)!$ in the denominator of
respectively the first and second above expressions are negative integer, they take infinite values which therefore truncate the infinite series.

\vspace{5 mm}
\textit{HOM Shot Noise calculation:}

Calculating the HOM noise of colliding electron pulses generated at opposite contacts when applying voltage $V_{L}=V(t-\tau/2)$ and
$V_{R}=V(t+\tau/2)$ on left and right contact respectively, is equivalent to calculate the noise when applying $V(t-\tau/2)-V(t+\tau/2)$ on the left
contact only. The resulting zero temperature shot noise is then:
 \begin{equation}\label{lesPIk}
    S_{I}/S_{I}^{0}= \sum_{k=-\infty}^{+\infty}|k| |\Pi_{k}|^{2}
\end{equation}
where
\begin{equation}\label{PIkdef}
    \Pi_{k}=\frac{1}{T}\int_{0}^{T} dt e^{-i\varphi(t-\tau/2)} e^{i\varphi(t+\tau/2)} e^{i2\pi k\nu t}
\end{equation}
\begin{equation}\label{PIk}
    \Pi_{k}=e^{-i2\pi\nu k \tau/2} (\sum_{l} p_{l} p_{l-k}^{\ast}e^{i2\pi\nu l \tau})
\end{equation}
where $p_{l}$ takes the same values as in the main text and is associated to the unshifted potential $V(t)$.
Because the effective voltage $V(t-\tau/2)-V(t+\tau/2)$ has symmetric weight for positive and negative values and is symmetric in $\tau$ we have
$\Pi_{k}=\Pi_{-k}^{\ast}$ . The shot noise computation then reduces to calculate:
\begin{equation}\label{SNPIk}
    S_{I}/S_{I}^{0}= 2 \sum_{k=1}^{\infty}|k| |\Pi_{k}|^{2}
\end{equation}
\vspace{5 mm}
\textit{HOM Noise of periodic single charge Lorentzian pulses:}

To calculate the $\Pi_{k}$ we use the exact values : $p_{(l<-1)}=0, $ $p_{-1}=-\beta$ and $p_{(l\geq0)}=(1-\beta^{2})\beta^{l}$
obtained for single charge $q=1$ Lorentzian voltage pulses
. Injecting these values in Eq.(\ref{PIk}) we find:
\begin{equation}\label{PI0}
    \Pi_{0}= \frac{1-(2-e^{i2\pi \tau})\beta^{2}}{1-\beta^{2}e^{i2\pi \tau}}
\end{equation}
\begin{equation}\label{PI02}
    \Pi_{l\geq1}= \beta^{l}(1-\beta^{2}) e^{-i2\pi l\tau/2} \frac{e^{i2\pi \tau} - 1}{1-\beta^{2}e^{i2\pi \tau}}
\end{equation}
This allows to calculate the HOM shot noise for single charge colliding periodic Lorentzian pulses:
\begin{equation}\label{SNLorPIk}
    S_{I}^{Coll}/S_{I}^{0}= (2 \sum_{l=1}^{\infty}l \beta^{2l})\left| \frac{e^{i2\pi \tau} - 1}{1-\beta^{2}e^{i2\pi \tau}}\right|^{2}(1-\beta^{2})^{2}
\end{equation}
which gives Eq.(\ref{HOMLorAnal}) of the main text.

To calculate, the temperature dependence one have to replace in Eq.(\ref{PI02}) $|l|$ by $l\coth(l h\nu/2k_{B}T_{e}) - 2k_{B}T_{e}/h\nu$.
Interestingly, as $\Pi_{l\geq1}=\Pi_{1}\beta^{2l}$, we see that the $\tau$ dependence is only included in $\Pi_{1}$. The temperature
dependence of the HOM shot noise decouples from the $\tau$ dependence. The temperature only reduces the noise but does not affect the shape.

\vspace{5 mm}
\textit{HOM Noise of periodic sine wave pulses:}

Rather than calculating the $\Pi_{l}$ using Eq.(\ref{PIk}), it is better to remark that the difference of two sine wave voltages, time shifted
by $\pm \tau/2$, is also a sine wave. We get $\Pi_{l}=J_{l}(2q\sin(\pi \nu \tau))$ for arbitrary charge $q$ per period. This gives
\begin{equation}\label{HOMSin1}
    S_{I}^{Col}/S_{I}^{0}= (2 \sum_{l=1}^{\infty}l J_{l}(2q\sin(\pi \nu \tau))^{2}
\end{equation}
and
\begin{equation}\label{HOMSin2}
    S_{I}^{Coll}/2S_{I}^{HBT}= \frac{\sum_{l=1}^{\infty}l J_{l}(2q\sin(\pi \nu \tau))^{2}}{2\sum_{l=1}^{\infty}l J_{l}(q)^{2}}
\end{equation}
The temperature dependence of the HOM signal can be calculated by replacing $|l|$ by $l\coth(l h\nu/2k_{B}T_{e}) - 2k_{B}T_{e}/h\nu$. We see however that the
series of Bessel functions in the above expressions does not allow a decoupling of the $\tau$ and $T_{e}$ dependence and the HOM shape is
expected to change with temperature. Indeed, the change affects electron and holes pairs with $lh\nu \lessapprox k_{B}T_{e}$ which interact with
thermal excitations.

\vspace{5 mm}
\textit{Wavepacket interpretation of the Floquet scattering approach:}

Here we show that the Floquet approach finds a useful wavepacket interpretation which allows to calculate the single electron wavefunction
in the time domain for single charge pulses. The wavefunction overlap can then be computed and be compared with
the HOM shot noise. This approach is restricted to zero temperature.

We start with the Martin Landauer wavepackets \cite{Martin92} defined in the energy bandwidth $h\nu$. Without loss of generality the wavefunctions
describing the states of the, say left, reservoir
$\sim e^{-i\varepsilon(t-x/v_{F})/\hbar}$ and labeled by the continuous energy variable $\epsilon$ can be transform into a set of orthogonal
wavepackets $\varphi_{n,l}(t-x/v_{F})$ defined by integration over the energy window $\varepsilon \in [l h\nu, (l+1) h\nu]$. Defining
$u=(t-x/v_{F})/T$:
\begin{equation}\label{WavePacks}
    \varphi_{n,l}(t-x/v_{F})=\frac{1}{\sqrt{2\pi \hbar V_{F}}}\frac{\sin(\pi (u-n))}{\pi(u-n)}e^{-i 2 \pi \nu(l+1/2) (u-n)}
\end{equation}
Using the annihilation operators $\widehat{a}_{n,l}^{0}$ acting on the Fock states of the unperturbed reservoir,
the Fermion operator before the action of the ac potential is:
\begin{equation}\label{WavePackPsi0}
  \widehat{\Psi}^{0} (t-x/v_{F})=\sum_{l} \sum_{n} \varphi_{n,l}(t-x/v_{F}) \widehat{a}_{n,l}^{0}
\end{equation}
with $\langle \widehat{a}_{n',l'}^{0 \dag} \widehat{a}_{n,l}^{0 } \rangle = \delta_{n,n'}\delta_{l,l'}f_{l}$ and $f_{l}=1$ for $l<0$
and $f_{l}=0$ for $l\geq0$. After the electrons have experienced the potential $V(t)$, the electron Fermion operator becomes:
\begin{equation}\label{WavePackPsi}
  \widehat{\Psi} (t-x/v_{F})=\sum_{k}\sum_{l} \sum_{n} p_{k} \varphi_{n,l}(t-x/v_{F}) \widehat{a}_{n,l-k}^{0}
\end{equation}
We see that the ac potential does not mix wavepackets of different $n$ but only wavepackets of different energies and same $n$ (we can also
see that $\sum_{l}p_{l}\varphi_{n,l}(t-x/v_{F})$ forms a new orthogonal basis of wavefunctions).

\vspace{5 mm}
\textit{Hong Ou Mandel correlation for single charge Lorentzian periodic pulses :}
We remind that the HOM noise associated with independent Lorentzian voltage pulses carrying one electron is directly related to the overlap of the electron wavefunction in the beam splitter. Here we would like to see if a  generalization is possible in the case of a periodic train of overlapping Lorentzian pulses still carrying only one electron. To do that, instead of the wave function overlap we consider the time average hermitian product of the electron Fermion operators with a relative time-shift $\tau$. Using the reduced variable $\theta=\tau/T$ it is given , before time averaging, by:
\begin{eqnarray}\label{WavePackOverLap}
  \langle \widehat{\Psi}^{\dag} (u+\frac{\theta}{2}) \widehat{\Psi}(u-\frac{\theta}{2})\rangle = \sum_{l l'} \sum_{n n'} \sum_{k k'} \nonumber \\
   p^{*}_{k'}p_{k} \varphi^{*}_{n',l'+k'}(u+\frac{\theta}{2})\varphi_{n,l+k}(u-\frac{\theta}{2})
  \langle \widehat{a}_{n',l'}^{0 \dag} \widehat{a}_{n,l}^{0} \rangle  & &
\end{eqnarray}

After time averaging and subtracting the contribution of the Fermi sea (i.e. with no
$V(t)$) we get :
\begin{equation}\label{WavePackOverLap2}
  \overline{\langle \widehat{\Psi}^{\dag} (u+\frac{\theta}{2}) \widehat{\Psi}(u-\frac{\theta}{2})\rangle } = \left(\frac{\sin(\pi \theta)}{\pi \theta}\right) ^{2} C(\tau)
\end{equation}

 where:
 \begin{equation}\label{CeTaCedur}
   C(\tau)=\left|\sum_{l}f_{l}e^{i2\pi l \theta}\sum_{k}(P_{k}-\delta_{k,0})e^{i2\pi k \theta}\right|^{2}
\end{equation}
For the presently considered case of periodic Lorentzian voltage pulses carrying unit charge and using the probabilities defined with respect to the total voltage $V(t)=V_{dc}+V_{ac}(t)$, i.e
$P_{0}-1=\beta ^{2}-1$ and $P_{l}=(1-\beta^{2})^2 \beta^{2(l-1)}$ for $l\geq1$, we get:
 \begin{equation}\label{CeTaCedurdur}
   C(\tau)=\frac{(1-\beta^{2})^{2}}{1-2\beta^{2}\cos(2\pi \tau/T)+\beta^{4}}
   \end{equation}
It is then straightforward to show that $1-C(\tau)$ yields the HOM shot noise given by Eq.(\ref{HOMLorAnal}), i.e. .
\begin{equation}\label{CeTaCedurdurdur}
   S_{I}^{HOM}=2S_{I}^{0} (1 - C(\tau))
   \end{equation}
Thus, even for overlapping pulses (but carrying a single electron), the HOM noise can still be interpreted as a measure of the overlap of the electronic wavefunction.
\begin{acknowledgments}

\end{acknowledgments}

\bibliography{References}

\begin{references}




\bibitem{Feve07} G F\`{e}ve, A. Mah\'{e}, J.-M. Berroir, T. Kontos, B. Pla\c{c}ais, D. C. Glattli, A. Cavanna, B. Etienne and
Y. Jin,  Science 316, 1169–1172 (2007)

\bibitem{Blumenthal07} M. D. Blumenthal, B. Kaestner, L. Li, S. Giblin, T. J. B. M.
Janssen, M. Pepper, D. Anderson, G. Jones, and D. A. Ritchie,
Nat. Phys. 3, 343 (2007)

\bibitem{Bocquillon12} E. Bocquillon, F.D. Parmentier, C. Grenier, J.-M. Berroir, P. Degiovanni, D.C. Glattli, B.
Pla\c{c}ais, A. Cavanna, Y. Jin, and G. F\`{e}ve,
Phys. Rev. Lett.   108, 196803 (2012)

\bibitem{Hermelin11} S. Hermelin et al. Nature 477, 437 (2011).

\bibitem{Mcneill11} R. P. G. McNeill et al. Nature 477, 439 (2011)

\bibitem{Splettstoesser08} J. Splettstoesser, S. Ol'khovskaya, M. Moskalets and M. Buttiker,
Phs. Rev. B 78, 205110 (2008).

\bibitem{Olkhovskya08} S. Olkhovskaya, J. Splettstoesser, M. Moskalets and M. Buttiker,
Phys. Rev. Lett. 101, 166802 (2008).

\bibitem{Splettstoesser09} Janine Splettstoesser, Michael Moskalets and Markus Buttiker,
Phys. Rev. Lett. 103, 076804 (2009).

\bibitem{Moskalets11} Michael Moskalets and Markus Buttiker
Phys. Rev. B 83, 035316 (2011)

\bibitem{Hack11} Geraldine Haack, Michael Moskalets, Janine Splettstoesser and Markus Buttiker
Phys. Rev. B 84, 081303 (2011).

\bibitem{Albert11} Mathias Albert, Christian Flindt and Markus Buttiker
Phys. Rev. Lett. 107, 086805 (2011).

\bibitem{Sherkunov12} Y. Sherkunov, N. d'Ambrumenil, P. Samuelsson, M. Buttiker,
Phys. Rev. B 85, 081108 (2012).

\bibitem{Jonckheere12} T. Jonckheere, J. Rech, C. Wahl, and T. Martin Phys. Rev. B 86, 125425 (2012)

\bibitem{coherentstates} L. S. Levitov, H. Lee, and G. Lesovik, J. Math. Phys. 37, 4845 (1996); D. A. Ivanov, H. W. Lee, and L. S. Levitov,
Phys. Rev B 56, 6839 (1997)

\bibitem{Keeling06} J. Keeling, I. Klich, and L. S. Levitov, Phys. Rev. Lett. 97, 116403 (2006)

\bibitem{Lesovik93} L. S. Levitov and G. B. Lesovik  , JETP
Lett. 5 8 , 230 ( 1993 ) 

\bibitem{Hassler08} see for example, F. Hassler, M. V. Suslov, G. M. Graf, M. V. Lebedev, G. B. Lesovik, and G. Blatter, Phys. Rev. B 78, 165330 (2008)

\bibitem{flying_quBits1} A. Bertoni, P. Bordone, R. Brunetti, C. Jacoboni, and
S. Reggiani, \emph{Phys. Rev. Lett.} \textbf{84}, 5912-5915 (2000).

\bibitem{flying_quBits2}  R.
Ionicioiu, G. Amaratunga, and F. Udrea, \emph{Int. J. Mod. Phys.}
\textbf{15}, 125-133 (2001).

\bibitem{flying_quBits3} T. M. Stace, C. H. W. Barnes, and G. J.
Milburn, \emph{Phys. Rev. Lett.} \textbf{93}, 126804-7 (2004).

\bibitem{Ji03} Y. Ji, Y. Chung, D. Sprinzak, M. Heiblum, D. Mahalu, and H.
Shtrikman, Nature London 422, 415 (2003)

\bibitem{Roulleau08} P. Roulleau, F. Portier, P. Roche, A. Cavanna, G. Faini, U.
Gennser, and D. Mailly, Phys. Rev. Lett. 100, 126802 (2008)

\bibitem{Geerligs92} L. J. Geerligs et al., Phys. Rev. Lett. 64, 2691 (1990).

\bibitem{Pothier92}  H. Pothier, P. Lafarge, C. Urbina, M. H. Devoret, Europhys. Lett. 17, 249 (1992).

\bibitem{Shilton96} J. M. Shilton et al, Journal of Physics: Cond. Matter 8, L531 (1996); V. I. Talyanskii et al. Phys. Rev. B 56, 15180-–15184 (1997).

\bibitem{Gabelli06} J. Gabelli,1 G. F\`{e}ve, J.-M. Berroir, B. Pla\c{c}ais, A. Cavanna,
B. Etienne, Y. Jin, and D. C. Glattli, Science 313, 499 (2006)

\bibitem{Hassler07} F. Hassler, G. B. Lesovik, and G. Blatter, Phys. Rev. Lett. 99, 076804 (2007)

\bibitem{Vanevic08} M. Vanevic, Y. V. Nazarov, and W. Belzig, Phys. Rev. B 78, 245308 (2008)

\bibitem{Lebedev11} A. V. Lebedev and G. Blatter, Phys. Rev. Lett. 107, 076803 (2011)

\bibitem{Haack10}  G. Haack, H. F\"{o}rster, and M. B\"{u}ttiker Phys. Rev. B 82, 155303 (2010)

\bibitem{Dubois12} J. Dubois, T. Jullien, P. Roulleau, F. Portier, P. Roche, Y. Jin, A. Cavanna, W. Wegscheider, and D.C. Glattli, submitted.


\bibitem{Moskalets02} M. Moskalets and M. B\"{u}ttiker, Phys. Rev. B 66, 205320 (2002)


\bibitem{Lee93} H. W. Lee and L. S. Levitov, arXiv:cond-mat / 9312013

\bibitem{Anderson67} P. W. Anderson, Phys. Rev. Lett., 18, 1049 (1967)

\bibitem{Lesovik89} G. B. Lesovik, Pis'ma Zh. Eksp. Teor. Fiz. \textbf{49} (1989)
513; JETP Lett. \textbf{49} (1989) 592.

\bibitem{Khlus87} V. A. Khlus, Zh. Eksp. Teor. Fiz. \textbf{93} (1987) 2179 [Sov.
Phys. JETP \textbf{66} (1987) 1243].

\bibitem{Yurke90} B. Yurke and G. P. Kochanski, Phys. Rev. B 41, 8184 (1990).

\bibitem{Buttiker90} M. B\"{u}ttiker, Phys. Rev. Lett. 65, 2901 (1990); Phys. Rev. B 46, 12485 (1992).

\bibitem{Martin92} Th. Martin and R. Landauer, Phys. Rev. B 45, 1742 (1992).

\bibitem{Blanter} Y. M. Blanter and M. B\"{u}ttiker, Phys. Rep. \textbf{336}, 1 (2000)

\bibitem{Lesovik94} G. B. Lesovik and L. S. Levitov, Phys. Rev. Lett. 72, 538 (1994).

\bibitem{Pedersen98} M. H. Pedersen and M. B\"{u}ttiker, Phys. Rev. B \textbf{58} (1998) 12993.

\bibitem{Rychkov05} V. S. Rychkov, M. L. Polianski, and M. B\"{u}ttiker, Phys. Rev. B 72, 155326 (2005)

\bibitem{Lee95} H. W. Lee and L. S. Levitov, arXiv:cond-mat/9507011

\bibitem{Schoelkopf98} R. J. Schoelkopf, A. A. Kozhevnikov, D. E. Prober, and M. J. Rooks, Phys. Rev. Lett. 80, 2437 (1998).

\bibitem{KozhevnikovThesis} A.A. Kozhevnikov's Thesis, Yale University (2001) http://www.yale.edu/proberlab/Papers/alexthesis.pdf

\bibitem{Shytov05} A. V. Shytov Phys. Rev. B 71, 085301 (2005)

\bibitem{Reydellet03} L.-H. Reydellet, P. Roche, D. C. Glattli, B. Etienne, and Y. Jin, Phys. Rev. Lett. 90, 176803 (2003)

\bibitem{Reznikov95} M. Reznikov, M. Heiblum, H. Shtrikman, and D. Mahalu, Phys. Rev. Lett. 75, 3340 (1995).

\bibitem{Kumar96} A. Kumar, L. Saminadayar, D. C. Glattli, Y. Jin, and B. Etienne, Phys. Rev. Lett. 76, 2778 (1996).

\bibitem{Pretre93} M. B\"{u}ttiker, A. Pr\^{e}tre, and H. Thomas, Phys. Rev. Lett. 70, 4114
(1993).

\bibitem{Moskalets06} M. Moskalets and M. B\"{u}ttiker, Phys. Rev. B 73, 125315 (2006)

\bibitem{Grenier12} C. Grenier et al., in preparation.

\bibitem{Grenier11} C. Grenier, R. Herv\'{e}, E. Bocquillon, F. D. Parmentier,
B. Pla\c{c}ais, J. M. Berroir, G. F\`{e}ve, and P. Degiovanni, New
Journal of Physics 13, 093007 (2011).


\bibitem{NoteBocquillon} In the different context of the periodic injection of energy resolved electron and holes above the Fermi sea from a
mesocopic capacitor, the noise of electron and hole partitioned by a QPC was found reduced by thermal excitations in ref \cite{Bocquillon12}
and its origin discussed and compared with experiments.

\bibitem{Torres01} J. Torr\`{e}s, T. Martin, and G. B. Lesovik, Phys. Rev. B 63, 134517 (2001).

\bibitem{Crepieux04} A. Cr\'{e}pieux, P. Devillard, and T. Martin, Phys. Rev. B 69, 205302 (2004).

\bibitem{Chevallier10} D. Chevallier, T. Jonckheere, E. Paladino, G. Falci, and T. Martin, Phys. Rev. B 81, 205411 (2010).

\bibitem{Jonckheere05} T. Jonckheere, M. Creux, and T. Martin Phys. Rev. B B 72, 205321 (2005).

\bibitem{Hong87} C. K. Hong, Z. Y. Ou, and L. Mandel Phys. Rev. Lett. 59, 2044 (1987)

\bibitem{Loudon89} H. Fearn and R. Loudon, J. Opt. Soc. Am. B, vol.6, 917 (1989)

\bibitem{Loudon98} R. Loudon, Phys. Rev. A 58, 4904 (1998)

\bibitem{Giavonnetti08} V. Giovannetti, D. Frustaglia, F. Taddei and R. Fazio, Phys. Rev. B 74, 115315 (2006)

\bibitem{Lebedev08} A. V. Lebedev and G. Blatter, Phys. Rev. B 77, 035301 (2008)

\bibitem{Glattli12} D. C. Glattli et al. in preparation.


\end{references}

\end{document}